\documentclass[12pt]{article}
\usepackage{graphicx}

\usepackage{latexsym}

\begin{document}
\title{Mass, Time, and Clock (Twin) Paradox in Relativity Theory}

\author{Anatoli Vankov,\\        
{\small \it  Bethany College, KS, anatolivankov@hotmail.com}} %,\\\
%{\small \it 16 Shawmut Avenue, Mansfield, MA 02048, USA}}

\date{}

\maketitle

\begin{abstract}
Historically, Einstein's Special Relativity Theory (SRT) came to reconcile the Galilean principle of relativity of inertial motion with the empirical fact of constancy of the speed of light in inertial coordinate systems. As a result, a classical (Galilean) coordinate transformation was replaced by the Lorentz transformation of 4-position and 4-momentum vectors in Minkowski space. This is the main content of SRT Kinematics, physical principles of which are clear and self-consistent. However, in many SRT applications questions arise concerning physical interpretation of geometrical structure of Minkowski space and proper/improper quantities resulted from Lorentz transformations. Often, poor or even wrong popular presentation of SRT concepts contribute to confusions and controversies especially when a problem formulation and/or its solution requires gaining an insight into theory physical foundations and beyond. The so-called clock (twin) paradox is an example of this kind of problem, disputed over years in literature; in different publications it was ill-posed in formulation, and/or ill-treated in solutions. The paper is intended to clarify operational meaning of mass and time quantities as main characteristics of an atomic clock, which is considered a quantum oscillator in association with the de Broglie wave concept. The specification of the concept of clock in quantum terms reflects the idea of relativistic mass and time complementarity, which is important for avoiding ambiguity of such notions as ``time rate'', ``time record'', and ``elapsed time'' under relativistic conditions. We used this approach in SRT Kinematics to conduct a detailed analysis of the clock paradox; results are discussed. It is also shown that in SRT Dynamics the proper mass must be acted by Minkowski force, what results in a clock rate variation.

 
\end{abstract}

%%%%%%%%%%%%%%%%%%%%%%%%%%%%%%%%%%%%%%

\section{Introduction}

Special Relativity Theory (SRT) in its basic part, Kinematics, was formulated by Albert Einstein in 1905 to reconcile Kinematics of Newtonian Mechanics with Electromagnetism in view of empirical evidence of constancy of the speed of light in all inertial frames. 
In both Newtonian and SRT Kinematics, the Galilean postulate of relativity of motion holds. However, SRT is a more general theory: given the postulate of the speed of light constancy, reduction of Lorentz to Galilean coordinate transformations is achieved with quadratic in $\beta$ terms neglected. Unlike Newtonian Mechanics,
SRT deals with point particles. However, it is not a theory shortcoming but rather its advantage because it enables us to recognize particle-wave duality of matter and establish SRT quantum connections by introducing a concept of atomic clock as a quantum oscillator. %An atomic clock in motion gives rise to the de Broglie waves and, correspondingly, space-time invariant scaling consistently with SRT postulates. 

Mathematically
SRT is a simple theory, but its consequences may be confusing for an {\em Euclidean mind}. There are numerous claims in ``non-mainstream'' literature that the postulates are wrong; ad hoc hypotheses are made 
admitting particle motion in Minkowski space faster than light. Empirical materials from time to time are presented allegedly revealing SRT inherent contradictions. For example, from observations of light from moving stars (or a radio echo from moving planets), it was found that the formula of velocity addition ($c_0\pm v$) rather than speed of light constancy fits the data. In this case, a confusion is caused by misunderstanding of SRT postulate of speed of light constancy. The speed can be measured in different inertial frames by the method of time of flight, the idea of which is, as follows. One should have a rigid movable frame with two thin transparent films, detectors of light, fixed at points $A$ and $B$ with a separation $d$. A light source should be attached to another movable frame.
The speed of light $c=d/t$ can be determined from measurements of a time interval $t=t_B-t_A$ where $t_A$ and $t_B$ are moments of light passages through detectors. Experiments show that the speed of light does not depend on whether the two frames are at rest or in a relative motion. The procedure of observations of light coming from a moving source is different; it is described by the Lorentz transformation of a 4-position vector of a photon moving in $x$-direction and having a coordinate $x=\gamma (c_0\pm v)t$, where $t$ is a time of light travel and $v$ is a speed of relative motion. The formula is, indeed, consistent with a classical photon model of aberration and Doppler effects (slow motion conditions) and observations. The addition ($c_0 + v$) in Lorenz transformations does not mean a motion with a speed exceeding the speed of light. Contrarily, it says that due to {\em the speed of light constancy}, the time of light propagation (before it hit a target) depends on a direction of relative motion: an event can be advanced or delayed. However, in classical mechanics, one may interprete the formula as if a superluminal motion is possible.

The addition ($c_0 \pm v$) seems to cause confusions as well in work\cite{1} devoted to ``reinterpretation'' of Einstein's illustration of relativity of simultaneity. At $c_0\to\infty$, time becomes the absolute Newtonian time and a simultaneity of two events becomes absolute, 
while in the SRT approximation of slow motion, predictions of all effects linear in $\beta$, {\em including the one of relativity of simultaneity}, remain consistent with observations to the precision of second and higher order terms. Thus, the idea of Einstein's imaginary train/embankment experiment is right and has a pedagogical value as far as it shows the relativity of simultaneity in prerelativistic (Galilean) Kinematics before actual introducing the relativity theory.

Under relativistic conditions of high-speed motion, another kind of controversies arises, first of all, in the connection with proper/improper categorization of physical quantities. This issue turned out to be the main problem in analysis of the so-called clock (twin) paradox widely known from fairy-tales about cosmic travelers getting younger by flying back and forth. 

%But could one be surprised to hear that everybody is practically involved in a relativistic experience of slow motion? Once in a while one may attampt to get through a glass door.
%bump into a glass door instead of getting through it. 
%While licking bruises, think about rightness of the relativity of motion and the related to it improper mass (energy of motion). Our comment could be helpful. The kinetic energy grows with   
%the squared speed (the Lorentz's $\gamma$ at small $\beta$ is a parabolic function and does not have a linear term). The good thing about bruises is that the quadratic term is accompanied by a factor ``half''. The $\gamma$ played a major role in the paradox origination when multiplied by an intangible time in space-time, but multiplied by a mass of your head, acquires practical significance...  

In the paper, much attention is paid to operational interpretations of SRT concepts in the imaginary experiment  methodolology around the main topic formulated in the title: mass, time, and the clock (twin) paradox in relativity theory. The concept of time in Minkowski space is closely related to the concept of atomic clock, a quantum oscillator, mysteriously driven by proper mass. One may consider a point particle in Minkowski space a quantum oscillator and a light emitter/detector. In physical reality, particles in motion by inertia do not exchange photons. Particle/particle interaction and particle/wave duality is subject to speculations in current field theories where concepts of real and virtual photons in connection with de Broglie waves are introduced. The present work is devoted to issues of SRT physical foundations, while a field theory problem is out of considerarion.
It follows from our discussions of {\it space-time and 4-momentum space complementarity} that the de Broglie wave concept is originated in SRT Kinematics and plays an important role in a clock (twin) paradox analysis in SRT Kinematics. An excursion into SRT Dynamics was also made with the conclusion consistent with observations that the proper mass should be subject to Minkowski force action resulting in a clock rate variation.

%\begin{equation}
%a=\left( 
%\begin{array}{rcl}
%x & y  \\
%1 & 2 
%\end{array}
%\right)
%\end{equation}

%\section{Special Relativity Kinematics}
\section{Point Particle in Minkowski Space}

\subsection{Lorentz transformations}

%\subsection{SRT Kinematics preliminaries }%and Lagrangian formulation of inertial motion problem}
%\subsection{Lorentz space-time coordinate transformation, and imaginary experiments}

%Special Relativity Mechanics is basically a theory of relativistic motion of a point charged particle acted on by an electromagnetic force in Minkowski space; so far, attempts to include other fundamental forces in a full range failed \cite{Goldstein}. Before going into this subject, we need to review principles of SRT Kinematics that is, the problem of free point particle motion in Minkowski space. 

Let us begin with known facts about SRT Kinematics. A free point particle in Minkowski space is characterized by a position 4-vector $X^\mu=(X^0,\ X^i)$, $\mu=0,\ 1,\ 2,\ 3$\ in an arbitrarily chosen coordinate system; $X^0=c_0 t$ is a time (temporal) part,  $X^i$,\ ($i=1,\ 2,\ 3$) are Cartesian 3-coordinates
representing a space (spatial) part of the vector;  $c_0$ is the speed of light in empty space. According to the SRT postulates, all inertial reference frames are equivalent, and the speed of light is constant in all of them. 
The position vector traces a trajectory of particle motion (a world line), which is a straight one.
In a rest frame, the vector takes the form $X^\mu=(c_0\tau,\ 0,\ 0,\ 0)$ where $\tau$ is a rest (proper) time distinct from  the so-called coordinate (or improper) time $t$. Thus, a world line of a rest particle coincides with $X^0$-axis. We shall see that for a free particle $\tau=t(1-\beta^2)^{1/2}=t/\gamma $. With the use of denotations of classical Cartesian coordinates $X^1=x ,\ X^2=y,\ X^3=z$, the position vector can be written $X^\mu=c_0 t(1, \ \beta^i)$, where $\beta^i=V^i/c_0$, and a 3-velocity vector has components $V^1=v_x=dx/dt$, $V^2=v_y=dy/dt$,  $V^3=v_z=dz/dt$. The inner scalar product is defined $X^\mu X_\mu=(c_0 t)^2-(x^2+y^2+z^2)$ (repeated upper and lower indeces summed up). It is said that the vector is constructed in the metric signature sign convention $(+,\ -,\ -,\ -)$, the quadratic metric form being $ds^2=dX^\mu dX_\mu=(c_0 dt)^2-(dx^2+dy^2+dz^2)=(c_0 dt)^2 (1-\beta^2)$, or $ds=c_0 d\tau$. The metric form is preserved (Lorentz invariant).  
The so-called proper 4-velocity $U^\mu=dX^\mu/ds$ is inttroduced, where $ds=(dX^\mu dX_\mu)^{1/2}$, and $U^\mu=(\gamma, \ \gamma \beta^i)$ is a tangent unit 4-vector with a squared length  $U^\mu U_\mu=1$. Wwith the above metric signature, vectors with a positive squared length are called time-like vectors. The 4-momentum vector for a point particle of a proper mass $m_0$ is defined $P^\mu=m_0 U^\mu$; its length (norm) is $m_0$.

%Having SRT posulates given and Minkowski space introduced, SRT Kinematics can be formulated in a relativistic Lagrangean framework (Landau \cite{Landau}).  
%With no external field, the Lagrangian can be chosen a constant, say, $L(s)=m_0$ in the action variation problem 
%\begin{equation}
%\delta S=m_0 \delta \int_a^b ds=m_0 \int_a^b d(\delta s)=0          
%\label{L7r}
%\end{equation}
%To set up the expression for $\delta S$, one has $ds=(dX_\mu dX^\mu)^{1/2}$ to substitute into (\ref{L7r}) $d(\delta s)=\delta (ds) =[\partial (ds)/\partial (dX_\mu)] \delta (dX_\mu)=(dx^\mu/ds) \delta(dX_\mu)=U_\mu d(\delta X_\mu)$. Upon integrating $m_0 \int_a^b U^\mu d(\delta X_\mu)$ by parts, one gets
%\%begin{equation}
%\delta S=m_0 U^\mu \delta X_\mu \Bigr|_a^b -m_0 \int_{a} ^{b} \delta X_\mu d U_\mu=-m_0 \int_{a} ^{b} \delta X^\mu (\partial U_\mu/\partial s)ds=0
%\label{L8r}
%\end{equation}
%with the condition $\delta X^\mu=0$ at fixed end points. From (\ref{L8r}), it follows $({\partial U_\mu}/{\partial s})=0$. Thus, a relativistic Lagrangean problem of a free particle in Minkowski space leads to a time-like trajectory of motion being a straight line, and the 4-momentum as well as 3-momentum vectors conserved. 

%\subsection{Lorentz transformations}

The question arises how to express an equation of particle inertial motion in different coordinate systems. One can consider a particle ``instantaneous''  image in two coordinate systems $\cal S$ and $\cal S'$ with axes $x,\ y,\ z$ \ parallel to $x',\ y', z'$, respectively, both systems being in relative motion in the $x$ direction with the speed $v=c_0 \beta$ ($\beta_x$-boost). Relativistic generalization of the Newtonian Kinematics is achieved when the Galilean transformation of classical coordinates in the considered case
%\begin{eqnarray}
$x'=x-vt,\ y'=y,\ z'=z,\  \ t'=t$
%\label{1}
%\end{eqnarray} 
is replaced by the corresponding (special) Lorentz transformation for space-time 4-coordinates
\begin{eqnarray}
x'=\gamma x-\gamma \beta (c_0 t),\ \ y'=y,\ z'=z,\ \ (c_0 t')=\gamma (c_0 t)-\gamma \beta x 
\label{2}
\end{eqnarray}
Both coordinate systems are chosen right-handed,  $\cal S'$  moves in  $\cal S$  in the positive $x$-direction, therefore,  $\cal S$ moves in  $\cal S'$  in the negative $x'$-direction. The inverse transformation is obtained by interchanging primed and unprimed coordinates and taking an opposite sign of $v$:    
\begin{eqnarray}
x=\gamma x'+\gamma \beta (c_0 t'),\ \ y=y',\ z=z',\ \ (c_0 t)=\gamma (c_0 t')+\gamma \beta x' 
\label{3}
\end{eqnarray}
%In terms of group theory methods, we deal with the Lorentz group where matrix operators describing coordinate transformations act on Minkowski 4-vectors ${\bf X}$. In our particular case of a ``$\beta_x$-boost'', the Lorentz operator ${\cal L}$ is
%\begin{eqnarray}
%{\cal L}=\left( 
%\begin{array}{ccp{.20in}c}
%\gamma & -\gamma\beta & 0 & 0 \\ 
%-\gamma\beta & \gamma & 0 & 0  \\ 
%0 & 0 & 1 & 0  \\
%0 & 0 & 0 & 1
%\end{array}
%\right), & \ {\bf X'}={\cal L} {\bf X}, & \ {\bf X}=({\cal L})^{-1} {\bf X'}
%\label{3a}
%\end{eqnarray}
%with Lorentz transformations, SRT Kinematics is formulated consistently with the Galilean principle of relativity of inertial motion and the postulate of speed of light constancy in all inertial frames. 
%It is customary to refer to imaginary observers with sets of ``standard'' clocks and rods when physical conditions and particle properties in two inertial coordinate systems are compared. 
There is the assumption in (\ref{2}), (\ref{3}) that coordinate origins $O$ and $O'$ coincide at $t=t'=0$  (a ``null'', or $0$-event, for short). So far, this concept looks vague. It will be carefully examined later, when a conceptual difference between $t$ and $\tau$ will be discussed. One could have already noticed that the $t$ in SRT plays the same role as that in Galilean Kinematics: multiplied by $v$, it gives a distance traveled by a particle. 
A relativistic equation of motion follows from the above transformations:
%\begin{eqnarray}
$x= v t$,  \  $x'=-v t'$. \
%\label{2a}
%\end{eqnarray}
%Mutually observed rescaling of space and time coordinates (relativistic effects) takes place because units of time and length became frame dependent. 
For an arbitrarily chosen direction of relative motion, there are so-called general Lorentz transformations, (\ref{2}), (\ref{3}) being a particular case of them. Regardless of type, Lorentz transformations establish one-to-one correspondence between two coordinate systems of inertial (that is, not accelerating and not rotating) reference frames.

%\subsection{Particle and light in Minkowski space, and complementarity concept}

Thus, any particle should be assigned 4-coordinates. Act of imaginary (nob-perturbing) observation of the particle at some 4-point is called ``an event''.
According to the Minkowski space concept, any event $P$ in a particle history is separated from the $0$-event by a 4-distance $OP$ such that the squared distance is a metric (quadratic) form invariant under the Lorentz transformation. Metric invariance can be visualized geometrically. Denote $\beta=\tanh \phi$, $\cosh \phi=\gamma$ and $\sinh \phi=\gamma\beta$. Then the transformation can be interpreted as a mapping 
$P$ in $\cal S$ onto $P'$ in $\cal S^\prime$: 
\begin{eqnarray}
x^\prime=x\cosh \phi-(c_0 t)\sinh \phi, \ & (c_0 t')=(c_0 t)\cosh \phi- x \sinh \phi 
\label{4a}
\end{eqnarray}
or, denoting coordinates $X^1=x$, $X^2=y$, $X^3=z$, $X^4=i c_0 t$ and defying $i \beta= \tan \psi$, where $\psi= {\it i} \phi$, ($\phi$ is purely real), $\cos \psi=\gamma$, \ $\sin \psi =i \gamma\beta$:
\begin{eqnarray}
{X'}^1=X^1\cos \psi+X^4\sin \psi, \ & {X'}^4=X^4\cos \psi- X^1 \sin \psi 
\label{5a}
\end{eqnarray}

It should be noted that in the above formalism, the metric sign convention is $(+,\ +,\ +,\ -)$, and the time-like vector has a negative squared length, for example, $(U^\mu U_\mu)^{1/2}=-1$. 
Formally, (\ref{5a}) describes a rotation of a complex plane $X^1,\ X^4$ at a purely imaginary angle $\psi (\beta)$. Thus, Lorentz invariance of Minkowski metric can be associated with a constant radius $OP$ of imaginary plane rotation.
Another picture of the Lorentz $\beta_x$-boost transformation can be drawn as a diagram in a real ($x,\ t$) plane showing primed and unprimed coordinate axes for different $\beta$ (sometimes it is called the Loedel diagram). Here, there is no room for those diagrams: they are given in many textbooks with explanations of geometrical treatments of relativistic effects, such as the time dilation and the length contraction. 
%One can relate relativistic effects to non-Euclidean space-time properties of high-speed world, where a motion of particles is relative but the speed of light is absolute in a certain sense. 
 
There is a difference in a pattern of motion of particles and photons. For particles, geometrical structure of Minkowski space is determined by the time-like base vector $U^\mu$ having the {\em proper} form $U_0=(1,\ 0,\ 0,\ 0)$ in a rest frame. In a coordinate system where the particle is in uniform motion in $x$-direction, the vector takes the {\em improper} form $U^\mu=(\gamma,\ \gamma\beta,\ 0, \  0)$. Of course, $\beta$-boost transformations can be successive. Figuratively, we deal with the SRT geometrical ``skeleton''  $U^\mu$ designed in accordance with the mathematical identity $\gamma^2-\gamma^2\beta^2=1$ (or equivalently $\cosh^2\theta-\sinh^2\theta=1$). The spatial part $U^i$ is related to the 3-velocity space described by the so-called Lobachevsky (hyperbolic) 3D geometry. In non-Euclidean terms, it is characterized by a constant negative curvature. Consequently, the parallel line axiom does not hold, and a classical velocity addition formula became invalid.
Physics begins with wrapping ``the skeleton'' in ``flesh'': one has to multiply the $U_0^\mu$ by the proper time interval $\Delta\tau_0$ or the proper (rest) mass $m_0$ (or some other proper scalars proportional to $\Delta\tau_0$ or $m_0$) to get space-time or 4-momentum space, respectively. Another base 4-vector  $u^\mu$ having an invariant null norm  $(u^\mu u_\mu)^{1/2} =0$ is needed to describe a light ray (a photon) propagation. For a photon in $(x, y)$ plane,  $u^\mu=(1,\ cos\theta,\ \sin\theta,\ 0)$. A photon ``skeleton'' $u^\mu$ is ``wrapped'' in a scalar ``flesh''  similarly to $U^\mu$. 

Two in a list of scalars seem to be special: mass and time. They give rise to 4-momentum and 4-coordinate space, which we incline to consider {\em complementary}. 
As is known, non-relativistic Quantum Mechanics provides the 3-coordinate and 3-momentum  complementary representation of objects (states). The complementarity concept plays an important role in explanation of Heisenberg uncertainty principle and, generally, Copenhagen school philosophy. In relativistic theories, 3-coordinate and 3-momentum measurements should be considered incomplete because of four dimensional nature of quantities, while measurements of 4-vector norms lead to determination of Plank constant. This follows from SRT quantum connections: $hf_0=m_0 c_0^2$, where $m_0$ is proper mass of a particle (an atomic clock). The corresponding proper frequency is proportional to the light frequency in the source frame $f_0=1/\Delta\tau_0$. A period $\Delta\tau_0=1/f_0$ can be considered a proper time interval related to the time-like metric $\Delta s_0=c_0 \Delta\tau_0$. Then a scalar product of vectors  $\Delta X^\mu$ and $P^\mu=m_0 U^\mu$ is a Lorentz invariant quantity related to physical constants:        
\begin{equation}
P^\mu \Delta X_\mu=h/c_0 
\label{6a}
\end{equation}
Let us introduce a proper time $\tau$ in the form $\tau=n \Delta \tau_0$, where $n$ is a number of de Broglie wave fronts in the world line interval. Then the corresponding improper time is $t=n \Delta t=n \gamma \Delta\tau_0 $. As far as $\tau^2$ is Lorentz invariant, the number $n$ (number of clock ticks) is Lorentz invariant as well. Indeed (putting for a while $c_0=1$, $h=1$): 
\begin{eqnarray}
X^\mu X_\mu=n^2 {\Delta \tau_0}^2 ,\ & P^\mu X_\mu=n 
\label{7a}
\end{eqnarray}
Those relations reflect {\em complenentarity} of 4-coordinate and 4-momentum vectors. 

The following comment may be helpful.  
In old terminology, a vector with an upper running index is a contravariant vector while the one with repeated lower index  is a covariant vector, or co-vector. Nowadays the latter in scalar products is often termed a 1-form (generally, N-form) studied in differential geometry. In our case, the 1-form is associated with the de Broglie wave: the dot-product gives a number indicating how many surfaces of equal phase are pierced by the vector (see, for example, Misner \cite{Misner}). In de Brogli wave terms, the number manifests Lorentz invariance of the phase difference between given end points in a particle world line. In physical applications, one should take into account that a co-vector 3-direction is opposite to that of a vector. 
%For a photon, $p^\mu \Delta x_\mu=0$ because a norm of both 4-position and momentum vectors is zero. However, $x^0x_0=x^i x_i=n^2 {\Delta x_0}^2 $, \ $p^0 \Delta x_0=p^i \Delta x_i=n$. Hence, complementarity takes place for temporal and spatial sub-spaces. 
An unalienable unity of mass and time concepts is seen from the fact that a frequency of both an atomic clock (a quantum oscillator) and an emitted photon is proportional to the mass of the quantum oscillator. 
%The complementarity concept will be discussed later in more details.  

\subsection{Treatment of proper/improper quantities in imaginary experiments}

With the help of ``imaginary observers'' we are going to explore operational meaning of Lorentz transformations. 
Let us consider Mary in $\cal S$ frame and John in $\cal S'$ both having standard clocks and rods.
In general, a quantum oscillator can play a role of standard (atomic) clock. It should function as a photon emitter/detector characterized by a specific electromagnetic transition between two quantum levels (see, for example, Moller \cite{Moller}). It could be a single particle or a collective atomic system of sufficiently small size, ideally, a point particle. An information exchange between observers means that a photon or a light ray emitted by one observer can be registered and analyzed by another observer. Sometimes, one can speak about making photographic light marks. The SRT concept of light (or photon) as an object in Minkowski space is a special issue, discussed later.

We are interested in relativistic effects such as the time dilation (retardation) and the corresponding length contraction. It was earlier emphasized that effects termed ``relativistic'' are those of second and higher order in $\beta$, as in $\gamma\approx (1+\beta^2/2)$ (appreciable when a particle speed is comparable with the speed of light). Technically, they are purely kinematical caused by the dependence of a time unit on a reference frame choice. Philosophically, they manifest a reconciliation of Galilean relativity principle with the postulate of $c_0$ constancy.  
The effects readily follow from Lorentz transformations. 

In the first imaginary experiment, John produced two short successive flashes at instants $t'=0$ and $t'=\Delta t'_0$ using his clock as a light emitter located at the origin $O'$. Mary was to detect the first signal at $t=0$ by her clock located at the origin (the $0$-event) and the second signal at $t=\Delta t$ by another clock at some point $x>0$ when the clock ``momentarily coincided'' with a location of John's moving coordinate origin $x'=0$ (the concept of ``clock coincidence'' will be discussed later). In accordance with (\ref{3}), Mary measured the improper time interval and compared it with the proper one: $\Delta t=\gamma \Delta t'_0$. Of course, conditions are symmetrical: John could detect Mary's similar signals and find $\Delta t'=\gamma \Delta t_0$ bearing in mind that $\Delta t_0=\Delta t'_0$ and $\Delta t=\Delta t'$ by definitions of standard clocks and measurement procedure. Thus, the improper time interval is always the factor $\gamma$ greater than the proper one. Though the term ``time dilation'' is often used, one compares standard clock rates rather than time elapsed.
 
In the second imaginary experiment, John placed a standard rod with its left and right ends at $x'=0$ and $x'=l'_0$, correspondingly, and put shining clocks there, signals from which Mary was supposed to detect at $t=0$ (simultaneously with the signal from the $0$-event). It is crucial in the length contraction concept that this is Mary, who detects simultaneously the two signals emitted by John not simultaneously.  
In accordance with (\ref{2}), the relationship $l=l'_0/\gamma $ takes place interpreted as the length contraction effect. Conversely, John could detect Mary's similar  signals and find $l'=l_0/\gamma $ bearing in mind that $l_0=l'_0$ and $l=l'$ due to the symmetry of measurement conditions. Thus, for any observer a moving rod appears shorter in the direction of motion. %Suppose, John triggered two flashes from $x'=0$ and $x'=l'_0$ simultaneously at $t'=0$. Then Mary, according to (\ref{3}), would find different result: $ l'_0=l/\gamma$. This example shows how definitions are important in interpretation of SRT results.

In practical (laboratory) experiments with high energy particles both kinematical relativistic effects are routinely observed and interpreted in terms of Lorentz transformations. For example, unstable energetic particles of a rest life-time $t_0$ travel a distance proportional to the time of flight $t=\gamma t_0$. 
Next we shall see how improper quantities appear in the 4-momentum space in connection with the photon exchange procedure.

%\section{4-Momentum (Complementary) Space and Relativistic Effects}
\section{Photon in Minkowski Space}

\subsection{Photon concept in SRT Kinematics}
%\subsection{Coordinate mapping using ``light marks''}

%A typical textbook explanation of how and why proper and improper time intervals originate is usually illustrated by a picture %$a)$ (Fig. \ref{Light path})
%of a light ray in ($x,\ y$) plane. The ray emitted in a $y$-direction from a light source (John's frame) moving in a positive $x$-direction appears to an observer at rest (Mary) to be ``dragged by inertia'', what makes a light travel distance greater than a vertical path in the light source (John's) frame. By virtue of a constancy of the speed of light in all inertial frames, it is concluded that a travel time in Mary's system is the factor $\gamma$ greater than in John's one: $ \Delta t =\gamma \Delta t_0$. This picture is intended to give the idea of the time dilation effect. Instead, it makes an attentive reader confused about the above formula obtained out of the SRT context. Since the Lorentz transformation is not considered there, the length contraction is not taken into account, and the concept of the speed of light constancy is abused. A classical photon is ``dragged'' as well because of the light aberration. This effect is linear in $\beta$ in a classical photon model obtained from SRT in a low-speed approximation. The latter means that quadratic $\beta$-terms responsible for the time dilation and length contraction  effects, are dropped. That is why the popular ``light travel picture'' allegedly illustrating relativistic effects is misleading. 

By applying the Lorentz transformation to photon vectors, one can learn more about the photon exchange procedure, the $0$-event concept, and relativistic effects. Let us again consider frames $\cal S $ (Mary) and $\cal S' $ (John) in the $\beta_x$-boost formalism. We assume that $x$ and $x'$ axes do not coincide but can be however close. (Further, small characters $u^\mu$, $x^\mu$ and $p^\mu$ are used in denotation of photon base vector, coordinates and momentum; as before, classical denotations of Cartesian coordinates are \ $x^1=x$,\ $x^2=y$,\ $x^3=z$,\ $x^0=c_0 t$). 
A photon emitted at $t=0$ from the origin $O$ in some direction $u^i$ is characterized by a 4-coordinate vector $x^\mu$:                                                         
\begin{eqnarray}
x^\mu=(c_0 t,\  x^i)=(c_0t) u^\mu,\ & u^\mu=(1,\  u^i),\ &   u^i u_i=1\ \ (i=1,\ 2,\ 3) %\ \Delta x^\mu=(c_0 \Delta t\ , \Delta x^i)=(c_0 \Delta t) u^\mu  
\label{8a}
\end{eqnarray} 
where $u^\mu=dx^\mu/c_0 dt$ is a photon base 4-vector 
%\begin{eqnarray}
%u^\mu=(1,\  u^i),\ &   \ u^i u_i=1 \ (i=1,\ 2,\ 3)
%\label{9a}
%\end{eqnarray} 
with spatial components $u^i$, directional cosines of light ray in the observer's system: 
$x=c_0 t\cos\theta$,\  $y=c_0 t\sin\theta$,\ $z=0$.  Because the speed of light is constant in all reference frames, the photon vector $u^\mu$ is a null 4-vector: $\ u^\mu u_\mu=0$; it determines the null invariant metric: 
%\begin{eqnarray}
$x^\mu x_\mu=0\ , \ (\Delta s)^2=\Delta x^\mu \Delta x_\mu=0$. 
%\label{10a}
%\end{eqnarray}
A radius $r$ of spherical light wave and its squared interval $(\Delta r)^2$ relate to the quadratic metric form 
$(\Delta s)^2 $:
\begin{eqnarray}
r^2=c_0^2 t^2=x^2+y^2+z^2  \ , &\ (\Delta s)^2=(\Delta t)^2-(\Delta r)^2=0 
\label{11a}
\end{eqnarray}
Thus, we have a light 3-sphere in a coordinate system where a light source is at rest (say, in John's frame); hence, a temporal component of ${x'}^\mu $ is the proper time $t'_0=t_0$:
\begin{eqnarray}   
{x'}^\mu=c_0  t_0 (1, \cos\theta',  \sin\theta',\ 0)  
\label{12a}
\end{eqnarray} 
The vector components in Mary's frame follow from the inverse Lorentz transformation 
\begin{eqnarray}  
t=\gamma t_0 (1+\beta \cos\theta')=t_0/\gamma(1-\beta \cos\theta)   
\label{13a}
\end{eqnarray}
\begin{eqnarray}
x=c_0 t \cos\theta=\gamma c_0 t_0 (\cos\theta'+\beta), & \ y=c_0 t \sin\theta=\gamma c_0 t_0 \sin\theta'   
\label{14a}
\end{eqnarray}
%and from (\ref{13a}), (\ref{14a})
%\begin{eqnarray}  
%t/t_0=1/\gamma(1-\beta \cos\theta)   
%\label{15a}
%\end{eqnarray}

From (\ref{13a}) and (\ref{14a}), important consequences follow.  The temporary component, when mixed with the spatial part in a Lorentz transform, becomes anisotropic: a light sphere in John's frame $r'=c_0 t_0$ takes a form of ellipsoid in Mary's frame. At $\cos\theta'=\pm 1 $ the addition formula takes place $x=\gamma (c_0\pm v) t$ or  $x=(c_0\pm v) t$  for a slow motion. (This issue was discussed in Introduction in connection with ``confusions''). At $\cos\theta'=0$, one gets the proper-improper time relationship $ t=\gamma t_0$. Hence, the time dilation effect arises when a photon is sent perpendicularly to the direction of motion. This result clarifies the concept of $0$-event and ``light photographic marks'' in the methodology of light signal exchange between observers (recall ``imaginary experiments'').

In accordance with (\ref{6a}) and (\ref{7a}), the improper time $t$ is a sum of $n$ wave periods related to atomic clock ticks: $t=n \Delta t$, where $ \Delta t=1/f$. Consequently, formulae (\ref{13a}-\ref{14a}) give us a description of light aberration and Doppler effects. A complete picture of relativistic effects will be revealed from a consideration of the problem in the momentum (complementary) space.

%We shall return to this issue when considering a photon presentation in the momentum space.
%There are conclusions drawn from the above results.
%1)  Considering a photon exchange between observers, one has to assume that $x$ and $x'$ axes in the $\beta_x$-boost cannot coincide, but could be however close. Then the time dilation effect follows from  the Lorentz transformation applied to the photon 4-coordinate vector. In terms of photon exchange procedure, the concept of $0$-event and ``light photographic marks'' is clarified: a light ray (photon) has to be emitted perpendicularly to the direction of relative motion, $\theta' =\pi/2$. 
%2)  As a result of relativistic light aberration, a temporal component, when observed from a frame in motion, takes an elliptic form (\ref{15a}). 

\subsection{Aberration and Doppler effect}

To derive formulae for aberration and Doppler effects, it is customary to consider a wave propagation in a moving refracting medium under assumption that the Huygens' principle is consistent with SRT, and the phase is invariant, Moller \cite{Moller}. Instead, we are going to take advantage of the complementarity concept by considering the Lorentz transformations  in the photon 4-momentum space. The $0$-event in this case means that tails of 4-momentum vectors in all reference frames are put in the same coordinate system origin. 
  
A photon 4-momentum vector can be presented in different forms. The 4-wave vector has momentum components divided by Plank constant $h$. At the same time, 
the photon frequency vector is proportional to the momentum one: $p^\mu=h f^\mu/c_0$ where $f^0$ is the standard (one may call it {\it proper}) frequency of the photon emitted by a standard atomic clock at rest; this frequency is proportional to energy of a quantum oscillator $m_0 c_0^2=hf_0$ in a rest frame. 
Obviously, if axes $x$ and $x'$ coincide, angles of emission/observation have only values $ \pm\pi $.
Let John's clock at $O'$ be an emitter of light of the proper frequency $f^\prime_0=f_0$ while Mary's clock at $O$ works as a spectrometric detector. 
John's frequency vector is 
%\begin{eqnarray}
$f'^\mu=(f_0 ,\ f_0\cos\theta^\prime ,\ f_0\sin\theta^\prime ,\ 0)$,  \
%\label{18}
%\end{eqnarray}
The inverse Lorentz transformation into Mary's coordinate system gives 
$f^\mu=(f,\ f\cos\theta,\ f\sin\theta,\ 0)$ with components 
\begin{eqnarray}
f=\gamma f_0+\gamma \beta  f_0 \cos\theta^\prime,\ f\cos{\theta}= \gamma f_0 \cos\theta^\prime+\gamma \beta  f_0 \ , &  f\sin{\theta}= f_0 \sin{\theta^\prime}   
\label{19}
\end{eqnarray}
%\begin{eqnarray}                                                                                           
%f\cos{\theta}= \gamma f_0 \cos\theta^\prime+\gamma \beta  f_0 \ , &  f\sin{\theta}= f_0 \sin{\theta^\prime}      
%\label{20}
%\end{eqnarray}

After simple algebra, we have                                                                                                                                                        
\begin{eqnarray}
f/f_0=1/\gamma \left(1-\beta \cos\theta\right) =\gamma \left(1+\beta \cos\theta^\prime\right)  
\label{22}
\end{eqnarray}                                                
\begin{eqnarray}
\tan{\theta^\prime}=\sin{\theta}/\gamma \left(\cos\theta-\beta\right),\ & \
\tan{\theta}=\sin{\theta^\prime}/\gamma \left(\cos\theta^\prime+\beta\right)  
\label{23}
\end{eqnarray}
\begin{eqnarray}
\cos\theta^{\prime} =(\cos\theta - \beta)/(1-\beta\cos\theta),   &  \cos\theta =(\cos\theta^\prime + \beta)/(1+\beta\cos\theta^\prime)
\label{23a}
\end{eqnarray}
The relativistic Doppler effect is given by (\ref{22}), while (\ref{23}), (\ref{23a}) are relativistic aberration formulae, which automatically take into account the Doppler effect. Recall that angles $\theta$ and $\theta^\prime $ are referred to the 3-vector $\bf u$ pointing at the same {\it out-going} photon in Mary's and John's frames, correspondingly. They are emission angles, interconnected by aberration formulae.  The observation angle  $\hat\theta=\theta-\pi$ is characterized by the opposite vector pointing at the {\it in-going} photon to be detected.   
The Doppler effect is usually presented in terms of 3-vectors $\bf w$ and $\bf u$: 
$f=f_0 /\gamma (1-{\bf w\cdot u}) $  
where ${\bf w\cdot u}=\beta \cos\theta$.

A process of relativistic effect formation can be visualized with the help of Fig. \ref{Ellipse}, where the Doppler effect graph for $\beta=0.7$ is plotted in polar coordinates. Because there is a symmetry of rotation of the plane about an $x$-axis, a wave-front circle, shown in the picture, is actually a 3-sphere in John's view, which appears to Mary as a stretched ellipsoid. There is also a left-right reflection symmetry with respect to the central ellipsoid section. A big section of the ellipsoid, the ellipse, is depicted  according to the equation $f=1/\gamma(1-\beta \cos\theta)$ with major axes $a=\gamma=1.4$, $b=f'=f_0=1$ and the eccentricity $\epsilon=\beta=0.7$; the radius $f(\cos\theta)$ emerges from the left focus $O$. This is how Mary observes a light wave from John's moving source. The unit radius-vector is directed at $\theta' $ with respect to the direction of motion (a positive $x$-direction), while its image is observed at an angle $\theta$. 

It should be noted that a pair of vectors to be compared $\overrightarrow {OP'}$ versus $\overrightarrow {OP}$,  are time component of the frequency/momentum 4-vectors in coordinate systems related to the source and the detector, correspondingly, while the transfer vector 
$\overrightarrow {P'P}$ is a linear combination of primed spatial components. 
From the picture, it is seen that the Doppler effect is formed in the process of a photon momentum transfer from one to another reference frame. 

There is a vector relationship between the observed photon momentum $|\overrightarrow {OP}|=f$ and the corresponding source momentum unit  $|\overrightarrow {OP'}|=f_0=1$. The vector addition diagram reads  $\overrightarrow {OP}= \overrightarrow {OP'}+ \overrightarrow {P'P}$ where  $|\overrightarrow {P'P}|=\Delta f$ is the momentum transferred in the direction of source motion $x$ in observer's coordinate system. As a consequence of (\ref{19}-{\ref{23a}), the transferred momentum $\Delta f$ as well as the observed momentum $f$ are linear functions of $\cos\theta'$    
\begin{eqnarray}
\Delta f=(\gamma-1)\cos\theta'+\gamma\beta, \  & \ f=\gamma +\gamma\beta\cos\theta'
\label{24}
\end{eqnarray}

% \begin{figure}
% \end{figure}

%\begin{figure}
%\includegraphics{fig_1}% Here is how to import PES art
%\caption{\label{fig:epsart} A figure caption. The figure captions are
%automatically numbered.}
%\end{figure}
%
%
\begin{figure}[t]
\includegraphics{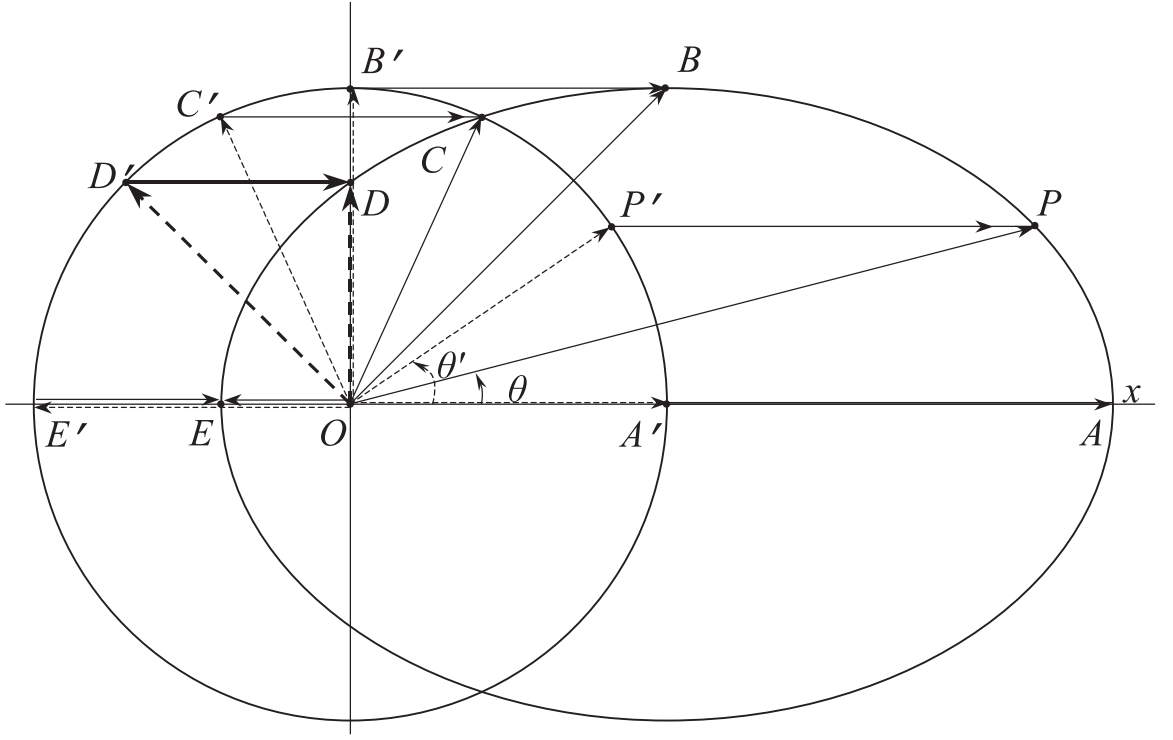}
%\label{Ellipse}
\caption{\label{Ellipse} Formation of Doppler effect and time dilation as transverse Doppler effect}
\end{figure}

%\begin{figure}
%\includegraphics{Anatoli_VankovFig1}
%\label{Ellipse}
%\caption{Ellipse and circle}
%\end{figure}

\subsection{Time dilation as transverse Doppler effect}

One can make a distinction between different stages in the Doppler effect observation when the source moved, first, towards the detector, and at some moment passed it by, and then kept flying away. Correspondingly, several specific cases are shown in the picture, one of them presents the time dilation effect. 

%At the first stage, according to (\ref{22}), John's signal was being received by Mary blue-shifted. At big distances, the emission angle in both systems, $\theta$ and $\theta^\prime$, was close to zero; consequently, the blue-shift effect was maximal, $f/f_0=\sqrt{(1+\beta)/(1-\beta)}$, then gradually decreased, and at some distance the frequency shift became zero. 

%With the angle $\theta$ rising further, the red-shift started to develop. At the moment when John passed Mary (that is, at the angle of observation $\pi /2$), the effect was $f/f_0=1/\gamma$. This is the transverse Doppler effect. When John kept flying away, the red shift progressed, gradually approaching the maximal (asymptotic) value 
%$f/f_0=\sqrt{(1-\beta)/(1+\beta)}$ at 
%$\theta\to 0$.
%  $\overrightarrow {OP}= \overrightarrow {OP'}+ \overrightarrow {P'P}$

{\it Case A}. \ At $\cos\theta^\prime =\cos\theta =1$ (an approaching source) the maximal blue-shift effect takes place $f=\gamma(1+\beta)=\sqrt{(1+\beta)/(1-\beta)}$;\ \ $\overrightarrow {OA}= \overrightarrow {OA'}+ \overrightarrow {A'A}$ ; \ $\Delta f=(\gamma -1)+\gamma\beta$. 

{\it Case B}. \  At $\cos\theta^\prime =0$,\ $\cos\theta =\beta$ (a ray perpendicular to the $x$-direction in John's frame), the blue-shift effect $f=\gamma$ results from \ $\overrightarrow {OB}= \overrightarrow {OB'}+ \overrightarrow {B'B}$; \ $\Delta f=\gamma\beta$. 

{\it Case C}. \ At $\cos\theta^\prime =-(\gamma-1)/\gamma\beta$,\ $\cos\theta =(\gamma-1)/\gamma\beta$\ , the Doppler effect is absent,  $f=1$. This results  from\ $\overrightarrow {OC}= \overrightarrow {OC'}+ \overrightarrow {C'C}$, \  $\Delta f=2(\gamma-1)/\gamma\beta) $. After this point, a range of the red-shift begins.

{\it Case D}\ (thick arrows). \ At $\cos\theta^\prime =-\beta$, $\cos\theta =0$\ (a ray perpendicular to the $x$-direction in Mary's's frame is detected). The transverse Doppler red-shift equivalent to the time-dilation effect,\ $f=1/\gamma$\ results from \ $\overrightarrow {OD}= \overrightarrow {OD'}+ \overrightarrow {D'D}$,\  $\Delta f=\beta $. Recall that the photon frequecy is proportional to the proper mass of emitter. After multilying the above vector relation by 
$\gamma^2 m_0^2$, we have the famous energy-momentum formula $E^2=\gamma^2 m_0^2=(\gamma\beta)^2 m_0^2+m_0^2$. 

{\it Case E}. \ At $\cos\theta^\prime =\cos\theta =-1$ (a source flying away) the maximal red-shift effect takes place,\ $f=\gamma(1-\beta)=\sqrt{(1-\beta)/(1+\beta)}$,\  $\overrightarrow {OE}= \overrightarrow {OE'}+ \overrightarrow {E'E}$ , \ $\Delta f=-(\gamma -1)+\gamma\beta$.

It is seen from the picture that for every ray of the directional cosine $\cos\theta^\prime$ in primed coordinate system another ray of the directional cosine  $\cos\theta$ exists in unprimed system such that $f(\cos\theta^\prime)\cdot f(\cos(\pi-\theta))=1$, as in A and E cases and B and D cases. For such cosine pairs, the identity $f_0=1/\Delta t_0$ and  $f(\cos\theta^\prime)\cdot f(\cos(\pi-\theta))=1$ takes place. The joining point is when $\cos\theta^\prime=-(\gamma-1)/\gamma\beta $ and $\cos\theta=(\gamma-1)/\gamma\beta $ resulting in $f=1$ (the Doppler shift is zero, case C). A complementary graph of a temporal component (\ref{13a}) will be similar to that of $f(\cos\theta)$, with the only difference that the origin should be placed to the other focal point. 
%The above scalar product is equivalent to $f^\mu  \Delta x_\mu=1$: this is how the complementarity concept works. 

%The following comment may be helpful.  
%In old terminology, a vector with an upper running index is a contravariant vector while the one with repeated lower index  is a covariant vector, or co-vector. Nowadays the latter in scalar products is often termed a 1-form (generally, N-form) studied in differential geometry. In our case, the 1-form is associated with the de Broglie wave: the dot-product gives a number indicating how many surfaces of equal phase are pierced by the vector (see, for example, Misner \cite{Misner}). In the de Brogli wave terms, the number manifests Lorentz invariance of the phase difference between given end points in a particle world line. In physical applications, one should take into account that a co-vector 3-direction is opposite to that of a vector. 
%Thus, $f^\mu \cdot \Delta x_\mu=n$ where $n$ is a real positive number such that $|x|=n|\Delta x|$. We shall see that this aspect is important for better understanding of the clock paradox. 

It is instructive to see how the relativistic picture looks in a low-speed approximation $\beta^2\to 0$, $\gamma =1$ (a classical photon model).
A classical Doppler effect along with light aberration is described by formulae    
\begin{eqnarray}
\Delta f=\beta, \  & \ f=1+\beta\cos\theta^\prime=1/(1-\beta\cos\theta)
\label{25}
\end{eqnarray}
\begin{eqnarray}
\cos\theta^\prime=\cos\theta/(1-\beta\cos\theta),   & \cos\theta=\cos\theta^\prime/(1+\beta\cos\theta^\prime),  \tan\theta=\tan\theta^\prime
\label{25a}
\end{eqnarray}
The ellipsoid becomes a sphere, and
the proper/improper time difference disappears. There is a price to be paid for accepting the approximation: in a classical photon model, one can distinguish between the effects caused by motion of source and detector. That means that the approximation breaks the symmetry of Galilean relative motion principle (in second order terms). In this situation, the principle should be rescued, for example, by means of an absolute ether of special properties. Historically, this idea was discarded by experiments, though some physicists keep advocating it in a broader (cosmological) sense. 

Nowadays, Einstein's SRT Kinematics is considered the only consistent theory of inertial motion world. 
An existence of ideal inertial frames is a special issue of Newtonian and relativistic Physics; it is discussed in literature, especially in cosmological theories. A cosmic background radiation seems to represent a preferred reference frame on a cosmological scale. In this connection, there are speculations about a relativity principle violation, which might appear in the form of ``forbidden by SRT'' frequency shift in electromagnetic radiation from distant stars or from relativistic particles accelerated in laboratories. So far, attempts to observe this kind of effect failed, for example, in a high-precision experiment \cite{Test1}, in which accelerated ions were used as a light source. The experiment showed that under special conditions when the relativistic Doppler effect was compensated, the time dilation effect automatically disappeared. At present, it is well understood that the time dilation is a part of the relativistic Doppler effect; most reliable time dilation tests were, in fact, high-precision Doppler effect measurements. 
An absolute impact of inertial motion on a time pace through time dilation is central in the clock paradox further discussed.

\section{Clock (Twin) Paradox}

\subsection{Clock paradox history and status}

In the clock (twin) paradox, the attention is focused on a possible impact of inertial motion on traveler's age. Suppose Mary and John were twins, and John went to a round trip while Mary stayed at home. What would be John's age in comparison with Mary's at the end of the trip? 
The paradox arose because the symmetry of observers' conditions was seemingly broken by the statement that John was in motion while Mary kept staying at rest. ``The round-trip condition'' is a cause of confusion. Upon John's return, clocks were brought to the same place for comparison, and it was believed that his wristwatch showed less amount of time than Mary's: as everybody knows, ``a moving clock runs slower''. %On the other hand, to make the problem formulation consistent with the Lorentz transformation formalism, each half of the imaginary trip should be started with $0$-event in order to treat the round trip as two independent successive one-way trips. Then both observers would have symmetric data and agree on the same travel time, but equal to what? 

It is not possible here to review the whole store of literature on the clock paradox; we can only draw a general picture of the paradox status development and confusions about it. 
Originally, the paradox was noted by Einstein who expressed the opinion that a traveling twin might return back to Earth ``younger''. Einstein left details of the problem to generations. In the centennial course of the paradox studies, the problem has been overgrown with numerous subtleties of minor or unknown importance.  
Meanwhile, experiments with relativistic short-lived particled confirmed the predicted time dilation effect in relativistic motion: they traveled longer distance than predicted by classical Physics. It was also evident that an acceleration did not spoil the effect, and it looks like a fast motion, indeed, ``prolongs'' a particle life.  A direct ``clock comparison'' (what is the essence of the paradox) cannot be done in this type of experiments; hence, they are not informative in regard to the paradox problem. %However, it is thought to be possible to interprete the results as if a meson decay rate might be slowed down by relativistic motion.

In another experiments, real atomic clocks were put into airplanes flying in different directions and eventually brought back home to be compared with ``a master clock''. An estimated difference in clock records at-rest versus in-flight
was tiny on a background of numerous dominant disturbances (gravitational force, Earth with atmosphere rotation, centripetal acceleration, and others). For a practical speed of plane, an expected relative effect would be of order $10^{-13}$, which is very hard to separate from an experimental noise and, what is even more important, from systematic errors due to model corrections. For this reason, trustworthiness of such measurements could not be high. So, we are left with logic of theory. 

Theoretical studies of the clock paradox have been conducted by many qualified physicists. In particular, the Doppler-shifted frequency as an additional means of the time rate counting along clock world lines was modeled. Some authors insisted that the problem had to be resolved only in the GRT framework to account for acceleration/deceleration stages (for example, Moller \cite{Moller}). However, it was generally realized that those effects can be eliminated (for example, Pauli \cite{Pauli}, recently Penrose \cite{Penrose}). 
During the 50s and 60s, especially intense disputes between prominent physicists took place over the paradox when two sides defended opposite views. Some scientists took a stand on a brink of denying the SRT validity (see original materials and further references in Tonnelat \cite{Tonnelat}, Sartori \cite{Sartori} and elsewhere, also Internet sites ``Twin (clock) paradox'' ).  
%How could several independent different treatments of the problem coexist (including the General Relativity framework claimed by authors to be the only appropriate for the problem)? When comparing proper and improper time counts by imaginary observers within Doppler effect approach, was a certain time period missing from being accounted? This problem was termed  ``the time gap'' explained by ``change observer's mind at the moment of changing a reference frame''. While all the above is referred to atomic clocks-particles, how do the results relate to aging of macrostructural bodies like human beings? 
From Penrose \cite{Penrose} one can infer that %those controversial questions are irrelevant because 
the problem 
should be treated in terms of matter properties governed by geometrical structure of Minkowski space, so the ``answer'' is readily found by inspection of clock world lines. This different viewpoint seems to be far from being satisfactory because it suggests that one should accept the proper/improper time relationship  $t=\gamma t_0$ as a key formula for a paradox solution without physical clarification.

%explanation of how to use it in the case of John's and Mary's world lines broken into pieces of several inertial stages. In mathematical terms, one deals with motion equations containing a jump discontinuity at end points of world line parts, and it is not clear how ``smoothing procedures'' could affect philosophy of clock comparison in the Lorentz transformation. 

A great variety of approaches to the paradox reflects differences (sometimes hidden) in its formulation. No wonder, different solutions may come out. 
At present, many physicists believe that the famous clock (twin) paradox academically does not exist, or it is ``apparent''. 
%A general opinion favored a statement that John's clock will indicate a lesser record of time passed as compared to Mary's. 
In fact, inconsistencies and contradictions in ``solutions'', subject to earlier criticism, were swept under a rug, while 
controversial statements about the paradox continue among those involved in physical research and educational activity. A historical and current importance of the paradox is evident from the fact that it raises questions related to SRT physical foundations and practical applications. Bearing this in mind, we are going to present our detailed analysis of the problem in parallel with criticism of its typical treatment.

\subsection{Time and time record}

In our view, controversies about the clock (twin) paradox basically arose because of ambiguity in terms  ``an elapsed time''  and ``a clock record of time'' treated equivalently in literature but actually having different physical meanings. Roughly speaking, suggested solutions reflected the idea that ``a moving clock runs slower than a rest one'': $t=\gamma t_0$. When a symmetry of relative motion is broken by John's forced turnaround, the time difference $t(Mary)- t_0 (John)$ seems to become absolute.  
The objection to this approach is that ``the moving, running slower clock'', that is showing the improper (``elapsed'') time $t$, does not exist in Nature: the $t$ quantity is a theoretical relativistic concept related to a set of clocks along a line of motion, as discussed next in details. 

Let us fix a {\it proper} distance $AB=x_0$ of John's travel with a relative speed $v$ from $A$ to $B$ under the assumption that shortly after he started and gained a constant speed, both observers zeroed their clocks (the $0$-event concept). By adopting philosophy of eliminating the role of end point neighborhoods in John's world line, we make the theory applicable. 
In Mary's (unprimed) coordinate system, the {\em improper}  time of one-way travel is $t=x_0/ v$. Because John observes the road of travel contracted, the {\it improper} length in his (primed) coordinate system is $x'=x_0/\gamma$ and the corresponding {\em proper} time $t'_0=x_0/\gamma v$. 
Observers must agree that measured proper/improper quantities are different purely because of a change of time/length units, but physically play the same role when enter a speed formula in the asymmetric problem formulation (Mary stays, John moves): $v=x_0/t=x'/t'_0 $.
%\begin{equation}
%\beta=x_0/t=x'/t'_0
%\label{26}
%\end{equation} 
If a symmetry of Mary's and John's conditions is admitted, we have an additional formula $v=x'_0/t'=x/t_0 $ (John stays, Mary moves).
%\begin{equation}
%\beta=x'_0/t'=x/t_0
%\label{27}
%\end{equation}
%A turnaround of one of the travelers does not make any difference. Once the role of end points was eliminated, the above SRT formulae remain valid: they manifest the Galilean principle of relativity of inertial motion, in accordance with the SRT postulates. 
Thus, the clock paradox resolution is a matter of proper/improper quantitiy interpetation at a conceptual level. 

In Relativity theory, the invariance of a world line length $ \tau $ between two points in Minkowski space is checked by the comparison of amount of clock ticks $n= \tau/\Delta \tau_0$, and $n=t/\Delta t$, where an atomic clock frequency $f_0=1/\Delta \tau_0$ is an inner particle property related to the proper mass. Thus, the number $n$ must be Lorentz invariant. How to fix end points and realize an information (photon) exchange between two observers, was previously explained. 
The basic idea of clock record of time was introduced  in  (\ref{6a}), (\ref{7a}), specifically, concepts of {\em ``clock record of time''} $n_0=f t=f_0 t_0$ and {\em ``elapsed time''} $t$. (We denoted the proper time unit $\Delta t_0$, which can be chosen arbitrarily; the same quantity will be denoted  $\Delta\tau_0$ when it is needed to emphasize that this is a particle inner property. So, $\Delta\tau_0\sim \Delta t_0$). It was also explained that more than one ckock is is needed to measure the impoper time. 
Consequently, in Lorentz transformations the improper time $t$, technically, is a cumulative sum of time intervals $t=\sum_{i=1}^{i=n} \Delta t_i $ where 
$ \Delta t_i =t(x_{i+1})-t(x_i)$ is a time difference shown by pairs of synchronized standard clocks at neighboring points $x_{i+1}$ and $x_i$ along the path $AB$ (in Mary's frame) in a process of tracking a single (John's) clock. Points $x_i$ are those, which were hit by John's light signals so that
measurements of $\Delta t_i$ is consistent with the procedure adopted for an $0$-event. It is appropriate to call the improper time $t$  {\em an elapsed (or cumulative) time}:  $t=n_0\Delta t$. John's wristwatch will show the corresponding cumulative (proper) time  $t'_0=n_0 \Delta t'_0$  along the same path $AB$. From the Lorentz transformation, it follows  $\Delta t'_0=\Delta t/\gamma$. The $\Delta t'_0$ determines
$n_0$ for a fixed interval $t_0$ if a proper frequency $f_0=1/\Delta t_0 $ is specified: $n_0=t/\Delta t=t_0/\Delta t_0$. 
What if a frequency changes due to acceleration during some time interval $\Delta T$? In this case, one should consider the integral $\Delta n =\int_{\Delta T} f(t) dt$ and a criterion $\Delta n/n$ of contribution of end point intervals to a clock record of travel time. A function $f(t)$ is an issue of SRT Dynamics. This is the number $n$, which is subject to comparison in the clock paradox. A comparison $t$ versus $t_0$ is not correct: when $\Delta n/n\to 0$, $(t-t_0)/ t_0$ may be huge.

\subsection{Clock comparison (examples)}

Let us apply the above concept of time record $n$ to the paradox problem. Further, we put $c_0=1$ for simplicity; a distance will be measured in ``light-seconds'', l$\cdot$s=$3\cdot 10^8$ $m/s$. It is also convenient to introduce two stationary observers, Mary at point $A$ and Lily at point $B$, with a proper distance $AB=x_0$ between them. The one-way cumulative time of John's travel in Mary's system is $t=x_0/\beta $ while his proper time is $t'_0=x_0/\gamma\beta $. 
There can be different variants of clock tick counts, when considering a world-line of John's motion.

{\em Variant 1: One-to-one clock comparison.}
In this variant, axes $x$ and $x'$ must coincide, and observers' wristwatch records are directly compared by counting light signals exchanged between the observers. A clock motion can be reversed, for example, by a means of elastic reflection at end points, and $0$-events can be determined there. In the first world-line leg of the John's trip $A\to B$, a red-shifted frequency $f_r=f_0\gamma(1-\beta)$ of John's signal was detected by Mary, a blue-shifted frequency  $f_b=f_0\gamma(1+\beta)$ by Lily, and conversely in the second leg $B\to A$. The frequency shift was caused exclusively by a longitudinal Doppler effect, while the proper frequency $f_0$ is related to the improper one $f=f_0/\gamma$\ (Case D in Fig. \ref{Ellipse}).
Mary's and Lily's estimate of John's cumulative (improper) time of travel was $t=n_0 \Delta t$ in each direction; it is related to the corresponding proper time $t'_0=n_0 \Delta t'_0 $ measured by John in each direction as well. (Recall, $\Delta t =\gamma\Delta t'_0$, \ $\Delta t'_0=\Delta t_0$, \ $f=1/\Delta t$). It should be emphasized that Mary measured the improper time of John's round trip $2t$ (but not $2t_0$) in spite of the fact that in this case she used her wristwatch but not a side clock off the $x$-axis. A total distance traveled by John includes two legs $A\to B$ and $B\to A$ with a $0$-event at each point. Thus, $AB=d_0=\beta t$, in accordance with the law of de Broglie wave propagation. Here $t=(t_B-t_A)$ and similarly for $BA$.  The proper distance $d_0$ was determined by the time of light flight method, in which a time of flight is also an improper time.

In the considered variant, the one-leg time of tick collection by a stationary observer differs from $t$: one needs to take into account a time delay of John's light signal. In addition to $t$, Mary had an extra time $x_0$ to allow the last red-shifted wave front (reflected from the point $B$) to reach her. Thus, the total time of Mary's collection of red-shifted photons was 
$t_r=x_0 +x_0/\beta=t(1+\beta)$. 
Contrarily, Lily (at point $B$) had to wait a period of time $x_0$ to allow the first blue-shifted wave front to reach her: $t_b =t(1-\beta)$. One can see that the time record of photon collection (the number of detected ticks) for both Mary and Lily is the same $f_r t_r=f t=n_0$ and $f_b t_b=f t=n_0$. The second leg of the trip was identical to the first one because observers just ``exchanged their names'' in counting procedures. {\em For both observers, the elapsed time of round-trip counting $t_r+t_b$ was continuous and equal to a total improper time $2 t$, a total time record being $2 n_0$}. 

What about John? He was engaged in similar to stationary observers' counting procedures: he received their blue and red shifted photons and made counts in a course of photon collection over times $t_r$ and $t_b$, correspondingly. In the first leg, he counted  Lily's ``blue photons'': $f_b t_b=\left[ f_0\gamma(1+\beta) \right ] \left[ t'(1-\beta) \right ]$, where $t'=\gamma t'_0$ is John's improper time of outer observer' motion. At the same time, he counted Mary's ``red photons'' with the same result: $f_b t_b=f_r t_r=f_0 t_0=n_0$. In the second leg, photons from Mary and Lily just ``exchanged color''; the result remained the same. Thus, everybody agreed on the clock record of time travel $n_0$ and the elapsed (improper) time of travel $t=x_0/\beta$, resulting numbers doubled for a round trip. The $t$ and $t_0$ quantities alone are not enough to compare twins' aging without a frequency analysis. Having the analysis done, twins would infer that their aging is not affected by a relative motion (provided, force effects were taken care of).

{\em Variant 2: One-to-many clock comparison} (axes $x$ and $x'$ do not coincide).
According to the Lorentz transformation scheme, in each leg Mary was supposed to detect the cumulative (improper) John's travel time $t$ using a set of side clocks along the 
$x$-axis: $t=n \Delta t_i$ is a collection of time differences calculated from records of neighboring clocks, as explained before. The detected light signals suffered a red shift due to the transverse Doppler effect independent of direction of motion. In this variant, the elapsed time of travel and the time of photon collection is the same quantity $t=x_0/\beta =n_0 \Delta t $ corresponding to the measured (improper) frequency $f=1/\Delta t $, \ $\Delta t =\gamma \Delta t_0$. A clock record of travel time would be $f t=n_0$. Lily had the same result as Mary due to symmetry of their conditions. 

On the other hand, John used his wristwatch to detect a cumulative (proper) time of travel in each leg 
$t'_0=n_0 \Delta t_0$. His clock record of one-way travel time was $ f_0 t'_0=n_0$ and doubled for the round trip. Observers eventually would meet each other having their life calendar intact in any variant (to the precision of acceleration effects, which can be eliminated). Next, we shall illustrate the clock paradox by graphical means as well.

\subsection{World line and de Broglie waves}

In connection with the clock paradox, it would be instructive to discuss geometrical images of the world line, as shown in 
three diagrams (Fig \ref{lines}) depicted for a positive $x$-direction, $0\le\beta <1$,\ ($c_0=1$). %\ because of shortage of space. 
%\begin{figure}
%\includegraphics{Anatoli_VankovFig4}
%\caption{\label{lines} 
%a) Could someone explain what ``light cone'' is in Minkowski space?
%b) World-line family of proper time count (4-distance) invariance
%c) World-line family of non-invariant 4-distance when distance of space travel fixed}
%\end{figure}
%Right-handed coordinate systems are used, and the Lorentz $\beta_x$-boost transformation is meant.
The diagram a) is known to illustrate  ``a world line'' in accordance with ``a light cone'' concept (as it is often presented in textbooks and SRT introductory lectures and papers). A vertical $x$-axis stands for a distance traveled by John in Mary's coordinate system. A horizontal $t$-axis is Mary's time such that $x=\beta t$, $\Delta x /\Delta t=\tan \alpha$. One can draw 
a triangle ${\cal A}_0{\cal B} {\cal A}_2$ (not shown in the picture) with vertices ${\cal A}_0 $ (start event), ${\cal B} $ (turnround), and ${\cal A}_2 $ (return event) with sides  ${\cal A}_0{\cal B}$ (outbound leg), ${\cal B} {\cal A}_2$  (inbound leg), and ${\cal A}_0 {\cal A}_2$ (Mary's waiting time). 
A general idea of ``proof'' that John's travel time record would be less than Mary's one is, as follows. %(Penrose \cite{Penrose}). 
In Euclidean geometry, side lengths satisfy $ {\cal A}_0{\cal B} +  {\cal B} {\cal A}_2  \ge {\cal A}_0 {\cal A}_2  $, while in Lorentzian geometry, all sides are time-like, and for this reason the inequality is reversed: $  {\cal A}_0{\cal B} +  {\cal B} {\cal A}_2  \le {\cal A}_0 {\cal A}_2 $. 

%\begin{figure}
%\includegraphics{Anatoli_VankovFig4}
%\caption{\label{lines} 
%a) ``Light cone'' diagram; \
%b) World-line family of invariant proper time; \
%c) World-line family of non-invariant proper time when 3-distance of travel fixed.}
%\end{figure}

In our view, the diagram a) is designed in a way making an analysis of the twin paradox and the above ``solution'' impossible, as explained next. Historically, the diagram was suggested to illustrate the SRT causality principle. One may consider ``ordering'' events in terms of past, present, future, and unreachable (``nothingness'') region outside ``light cone'' (Synge \cite{Synge} and elsewhere). 
%For example, let us choose a midpoint ${\cal A}_1 $ (turnround) in Mary's line ${\cal A}_0 {\cal A}_2= {\cal A}_0 {\cal A}_1 +  {\cal A}_0 {\cal A}_2 $ and put it in the coordinate origin. 
%Then, all world lines passing through the ``present'' event ${\cal A}_1 $ will fill space-time in such a way that the whole space-time is divided into three parts consisting of events {\em before} (region $B$), {\em after} ${\cal A}_1 $, (region $A$), and also unreachable events (region $U$, sometimes called ``nothingness''). A 3D sheet dividing $B$ and $A$ from $U$ is a light cone. One can obtain a 
The Euclidean picture can be obtained by ``opening out'' the light cone starting from $\alpha=\pi /4$ until it becomes flat at $\alpha=\pi/2$. In this conversion of 4-space into 3D space with the absolute time as a parameter, the above ordering of the events is preserved in Newtonian Physics. The diagram has a sense in illustration of the causality problem but
%The drawback of the diagram is that having $\tan\alpha =\beta$, 
it cannot show, in principle, how the Minkowski metric works and the Lorentz $\gamma$-factor originates.  
%It makes it irrelevant to our problem. 

Next diagrams b) and c) are suited for discussions of the paradox. A vertical $x$-axis indicates a proper distance (in light-second units) traveled by a particle and represents a spatial dimension in Minkowski space. A horizontal $\tau$-axis stands for a proper time (previously, a denotation $t_0$ was used). A family of projected lines $t(\tau)=\gamma \tau$ are shown having a slope $\tan\alpha=\gamma\beta$. The time-like metric is seen from the picture: $\Delta\tau^2=\Delta t^2-\Delta x^2$, that is $\tau=t/\gamma$, where $\sin\alpha=\beta$,\ $\cos\alpha=1/\gamma$. By this way, we present Minkowski geometrical objects. The diagrams have a reflection symmetry for negative values of $x$ in accordance with philosophy of symmetry of direct and inverse Lorentz transformations.

\begin{figure}
\includegraphics{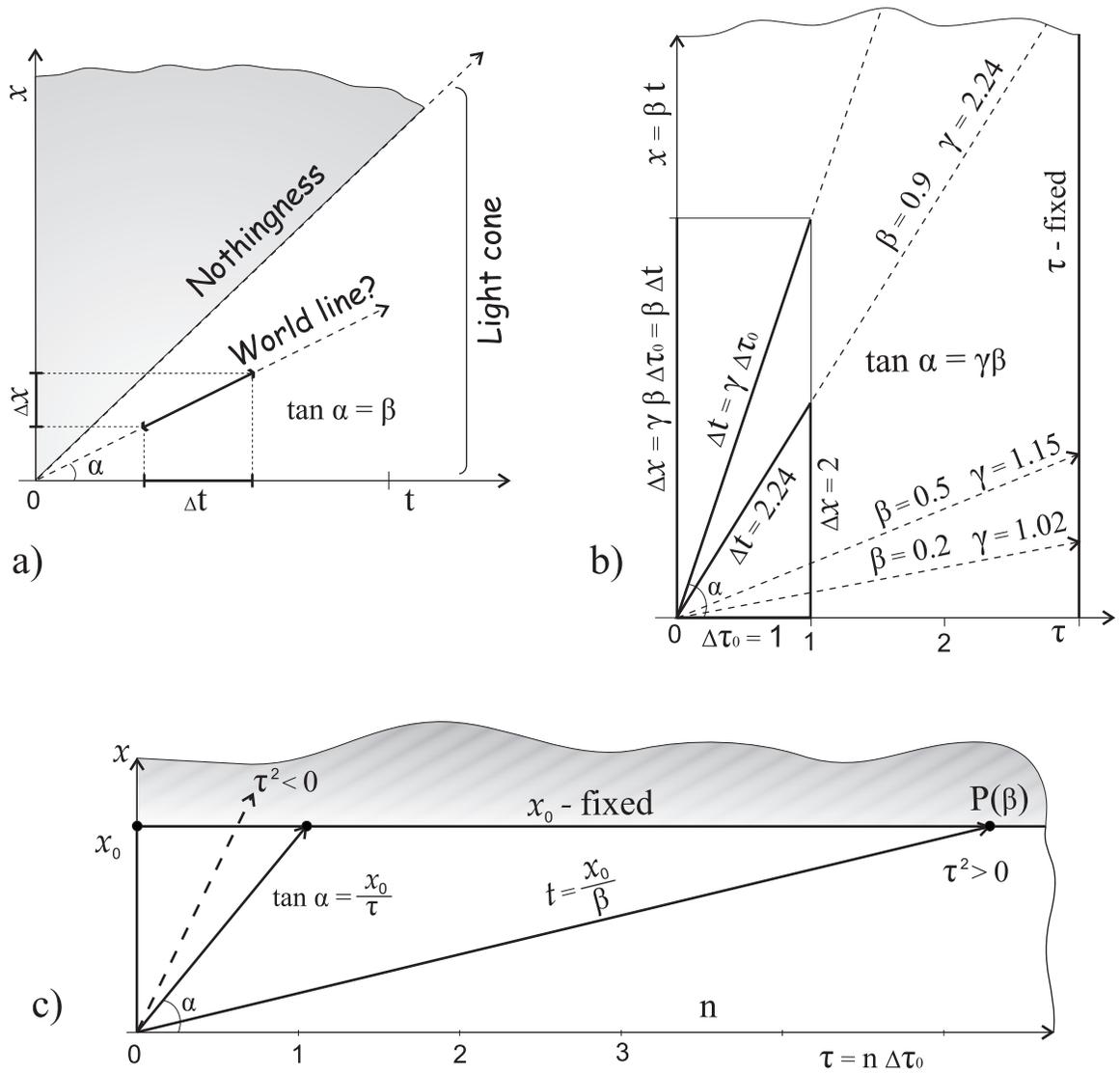}
\caption{\label{lines} 
a) ``Light cone'' diagram; \
b) World-line family of invariant proper time; \
c) World-line family of non-invariant proper time when 3-distance of travel fixed.}
\end{figure}

The diagram b) shows a world-line family obeying a constraint $\tau=const$ with $\beta$ as a parameter. In other words, the picture manifests the Lorentz invariance: a tail of each vector originates at a common $0$-point and a tip hits a point on a vertical line $\tau =\tau_0=n_0\Delta\tau_0$ with the clock record 
$n_0=\tau_0/\Delta\tau_0$. There are geometrical relations $\tau_0/t=\cos\alpha$,\ $x/t=\sin\alpha$,\ $\tau_0^2=t^2-x^2$ and the working formula $x=\gamma\beta\tau_0$ with SRT quantities $t=\gamma\tau_0$, $x=\beta t$. The diagram illustrates the clock record concept for the following clock paradox formulation: {\em given a fixed proper time, find a family of lines characterized by the invariant clock record $n_0$ for different $\beta$ and, correspondingly, cumulative time $t$. A method of exclusion of end point intervals is advised. It is understood that a round trip is equivalent to combined two one-way trips in opposite directions}.

Let us consider a particle having a rest life-time $\tau_0 $ in an example of $\beta=0.895\approx 0.9$ \ ($\gamma=2.24$, $\gamma\beta=2$)\  shown in the picture. In a laboratory coordinate system, the particle travels a proper distance 
$x_0=\beta t$, where $t=t_B-t_A=\gamma\tau_0$ is the time of flight. We choose the proper time interval $\Delta\tau_0=1$ \ $s$.
For a muon with a rest life-time $\tau_0=2.2\cdot 10^{-6}$\ $s$, a distance of particle travel is about $x=2.0\cdot10^{-6}$\ $l\cdot$s, $x_0=4.4\cdot 10^{-6}$\ $l\cdot$s, and $n_0=2.2\cdot 10^{-6}$\ ticks. A distance $x_0$ and, correspondingly, a time $t$ rises with $\gamma$ without limit when $\beta\to 1$, while $n_0$ is conserved. For example, ultra-relativistic muons, existence of which in primary cosmic rays cannot be excluded, will travel huge distances in spite of their short life-time. For energy of about $10^{21}$ $eV$ 
($\gamma=10^{13}$), a muon would travel a distance about one light-year before decay.    
One can arrange an elastic collision to reflect a relativistic particle from a midpoint of a traveled distance back to its source, as in twin paradox. 
According to our analysis, stationary and traveling clocks will indicate similar records of travel time $n_0$, no matter how different are $t$ and $\tau$, as was argued earlier. 

The diagram c) is suited for another formulation of the twin paradox relevant to discussions of the clock record concept in the paragraph 
`` One-to-many clock comparison'': {\em given a fixed (proper) distance $x_0$ of space travel, show a family of world lines with the time travel as a function of $\beta$.} Due to the constraint, the  4-vector family does not preserve the proper time $\tau$  or the clock record ticks $n$. Of course, a symmetry of direct-inverse Lorentz transformations (an equivalence of Mary's and John's points of view) remains intact. There are geometrical relations $\tau/t=\cos\alpha$,\ $x_0=t\sin\alpha$,\ $x_0^2= t^2- \tau^2$ and a working formula $n=x_0/\beta\gamma\Delta\tau_0=x_0/\beta t$. It is seen that a particle speed cannot exceed the speed of light. From an obsever's point of view, a cumulative (improper) time of John's flight in Mary's coordinate system $t=x_0/\beta$ cannot be less than a time of light flight $x_0$. The proper (John's) time of travel $\tau=t/\gamma =t\cos\alpha $ and the corresponding number of ticks $n$ can be however small, when $\beta\to 1$. In this limit, a particle vector $\overrightarrow {OP}$ becomes however close to the null vector $\overrightarrow {Ox_0}$. The latter cannot be extended into a darkened region $x>x_0$ because it would mean a motion with the speed greater than the speed of light. The diagram does not give a consistent presentation of vectors with $\tau<0$; for this reason, the region $x>x_0$ should be out of consideration. 
The following numerical examples for $\Delta\tau_0$= 1 $s$,  $x_0$= 1  $l\cdot$s for different $\beta$ are helpful.   
For $\beta=0.1$, $t\approx\tau \approx 10$ $s$, $\tau=n=10$ $s$\ (tick) (classical approximation: $\gamma\approx 1$,  $t\approx\tau \approx x_0/\beta $ . For $\beta=1/\sqrt 2$\ ($\gamma\beta=1$): $t=1.41$ $s$, $\tau=n=1$ $s$\ (tick) (relativistic motion). For $\gamma=10$: $t\approx 1$ $s$, $\tau=n=0.1$ $s$\ (tick); for $\gamma=100$: $t\approx 1$ $s$, $\tau=n=0.01$ $s$\ (tick), and so forth with  $\tau=n=1/\gamma$ (ultra-relativistic motion).

The fact, which cannot be realized by Euclidean mind, is that a particle with a however small life-time can travel a however great distance. The non-Euclidean explanation comes from the de Broglie wave phenomenon, which should be considered in SRT as an additional postulate of phase invariance of the de Broglie wave.
Consider the 4-wave vector in the $x$-direction. It is proportional to the 4-momentum (frequency) 
$f_0$ $(\gamma,\gamma\beta,\ 0,\ 0)$, so its velocity dependent components are $\omega\sim\gamma$, $k\sim\gamma\beta$. If the phase difference of a 4-wave $(\omega t-k x)$ is preserved, one can derive phase and group velocities: 
$\beta_{ph}=x/t=\omega/k=1/\beta$, and $\beta=\Delta x/\Delta t$.

During an elapsed time $t=x_0/\beta$ of travel   
over a fixed distance $x_0$, the wave will produced $n$ oscillations (ticks) at $n$ points $x_i$, $n=t/\Delta t$. The question arises: which is the standard rod to measure the distance 
$x_0$ with?  Because light and particle waves propagate out of phase, a proper length unit must be independent of atomic clock frequency. Nature gives us a natural all-wave length unit, a distance $AB=\Delta x_0$ traveled by light per a time unit ${\Delta t}^*=\tau_B-\tau_A$, as measured by the time of flight method. Thus, in metric determination and metrology standartization we have a standard time unit $\Delta\tau_0$ (from the atomic clock relationship $m_0 c_0^2=h/\Delta\tau_0$) and a standard length  $\Delta x_0$ (from the speed of light constancy 
$c_0=\Delta x_0/{\Delta t}^*$). The latter is defined in a description of time of flight measurements with photons of {\em arbitrary} energy frequency. In SI, the second is a duration of 9 192 631 770 periods of a radiation (standard photon) transition between two hyperfine levels of the ground state of the cesium 133 atom; the meter is the length of the path traveled by light in vacuum during a time interval of 1/299 792 458 of a second. Among all photons, let us chose a standard photon, for which a relationship ${\Delta t}^*=\Delta t=\gamma\Delta\tau_0 $ takes place. Then we connect both units by a means of introduction of the improper time interval concept $\Delta t $. To measure the $t$, more than one clock are needed. Eventually, we have the correlated time and length units related to universal constants $c_0$, $m_0$, $h$. 
An operational meaning of both units were illustrated by diagrams b) and c).
One can derive the relationship of the de Broglie wave length $\lambda_{dB}=h/c_0 m_0\gamma\beta$ and the standard length $x_0$. In a particular case $x_0=\lambda_0=c_0\Delta\tau_0=h/m_0c_0$, we have $\lambda_{dB}/\lambda_0=n$. This is a number of ticks in diagram c).

\section{From Kinematics to Dynamics}

A close relationship between mass and time in SRT Kinematics is expected in SRT Dynamics in connection with Minkowski force acting on the proper mass and by this way influencing the clock rate consistently with observations. The Minkowski force can be generally defined
\begin{equation} 
K^\mu=dP^\mu/ds=U^{\mu}(dm/ds)+m(dU^{\mu}/ds) 
\label{1r}
\end{equation}
In the conventional Relativistic Mechanics, the generalized Newton's law has a form 
\begin{equation} 
dP^{\mu}/ds=m dU^\mu /ds= K^\mu 
\label{2r}
\end{equation}
with the constant (invariant) proper mass, that is $dm/ds=0$. A proper mass variability in a force field was discussed in literature (for example, Moller\cite{Moller}, Synge\cite{Synge}, with references), but its consequences were not thoroughly realized yet. The speculative (alternative) approach to SRT Dynamics and issues of its empirical verification were studied by Author\cite{Van1},\cite{Van2}. Next, we are going to show that a field-dependent proper mass can be introduced in the relativistic Lagrangean framework in a general form. In order to derive covariant equations of motion, we apply the Hamilton's extremal action principle with a proper Lagrangian (in the case of a static potential field) 
\begin{equation} 
L(s)=-m(s)-W(s)
\label{3r}
\end{equation}
where $s=s(x^\mu)$ is a world line (arc)length, and a field is characterized by potential energy $W$ measured by a test point particle so that $W(x^\mu)\to 0$ at $x^\mu \to \infty$ (an equivalent mass-energy unit is used for convenience). Thus, 3-velocity $\beta$ does not appear in the Lagrangian. The variable proper mass in the Lagrangian describes a kinetic energy formation: its change relates to a potential energy change under a force field action. We want to show a consistence of the Lagrangean method with (\ref{1r}) and (\ref{3r}). The question arises: how does one know whether the proper mass is constant (as assumed in current field theories) or field dependent (as suggested here)? Our viewpoint is that the proper mass constancy is the assumption which is neither justified by direct experiments nor follows from first physical principles: this is the issue of theory physical foundations and subject to experimental falsification for every fundamental force.

The equations of motion in a covariant form are obtained by applying the Hamilton's Principle. The world line length is varied between two fixed end points $a$ and $b$:   
\begin{equation}
\delta S=\delta \int_{a}^{b}L(s)ds=\int_{a}^{b}\delta L(s)ds=\int_{a}^{b} \delta s \left({\partial L(s)}/{\partial s}\right) ds=0          
\label{4r}
\end{equation}
It should be noted that improper time $t$ does not appear at this stage because the problem is formulated in a rest frame.
We look for equations of motion giving the solution $x^\mu(s)$ and $m(s)$; consequently, (\ref{4r}) is taken in the form
\begin{eqnarray}
\delta S= \int_{a} ^{b} \frac{\partial L(s)}{\partial s} \delta s\ ds=\int_{a} ^{b} \delta s\ dL(s)=0
\label{5r}
\end{eqnarray}
The action variation (\ref{5r}) is identically equal to zero if $L(s)=Const$; this is the expression of conservative properties of field, which will follow from the equations of motion, if the temporal component of Minkowski force is put to zero. To get the equations of motion in an explicit form, one has to conduct the integration by parts bearing in mind that $\delta s =0$ at the endpoints $s=a$ and $s=b$, and differential and variational operators $d$ and $\delta$ are commutative:
\begin{equation}
\delta S=L(s)\delta s \Bigr|_a^b -\int_{a} ^{b}L(s)d(\delta s)=-\int_{a} ^{b}L(s)d(\delta s)=0
\label{6r}
\end{equation}
Before proceeding further, let us consider a particular case of a free particle motion discussed in Landau \cite {Landau} (Chapter 2), with the Lagrangian $L(s)=-m_0$ (in our denotations) and the action variation 
\begin{equation}
\delta S=m_0 \delta \int_a^b ds=m_0 \int_a^b d(\delta s)=0          
\label{7r}
\end{equation}
To set up the expression for $\delta S$, one has to start with $ds=(dX_\mu dX^\mu)^{1/2}$, substituting into (\ref{7r}) $d(\delta s)=\delta ds =(\partial ds/\partial X^\mu) \delta (dX^\mu)=(dX_\mu/ds) \delta(dX^\mu)=U_\mu d(\delta X^\mu)$ with $U_\mu(s)=dX_\mu/ds$ and integrating $m_0 \int_a^b U_\mu d(\delta X^\mu)$ by parts
\begin{equation}
\delta S=m_0 U_\mu \delta X^\mu \Bigr|_a^b -m_0 \int_{a} ^{b} \delta X^\mu d U_\mu=-m_0 \int_{a} ^{b} \delta X^\mu (\partial U_\mu/\partial s)ds=0
\label{8r}
\end{equation}
Again, we used a condition $\delta X^\mu=0$ at fixed end points. From (\ref{8r}), it follows for a free particle that $({\partial U_\mu}/{\partial s})=0$ that is, trajectories of free particles in the Minkowski space are straight lines, and the 4-momentum is conserved. 

Back to the general case: substituting the expression $d(\delta s)=U_\mu d(\delta X^\mu)$ into (\ref{6r}) and integrating by parts gives
\begin{equation}
\delta S=L(s)U_\mu \delta X^\mu \Bigr|_a^b -\int_{a} ^{b} \delta X^\mu d\left[L(s)u_\mu(s)\right]=-\int_{a} ^{b} \delta X^\mu d\left[L(s)U_\mu(s)\right]=0
\label{9r}
\end{equation}
or
\begin{equation}
\delta S=-\int_{a} ^{b} \delta X^\mu \frac{\partial \left[L(s)U_\mu(s)\right]}{\partial s} ds=0
\label{10r}
\end{equation}
Because variations $\delta X^\mu$ are independent for different $\mu$, the equality $\delta S=0$ in (\ref{10r}) is possible if and only if 
\begin{equation}
\frac{\partial \left[L(s)U_\mu(s)\right]}{\partial s}=0
\label{11r}
\end{equation}
Finally, with the Lagrangian (\ref{3r}) substituted into (\ref{11r}), we have the equations of motion
\begin{equation}
\frac{\partial \left[m(s)U^\mu(s)\right]}{\partial s}=-\frac{\partial \left[W(s)U^\mu(s)\right]}{\partial s}
\label{12r}
\end{equation}
where the index $\mu$ is raised for a presentation convenience.
Together with the independent equation 
\begin{eqnarray}
U_\mu U^\mu=1, \ & \ U_\mu (dU^\mu/ds)=0
\label{13r}
\end{eqnarray} 
characterizing the time-like character of massive particles, they allow us to determine five quantities $x^\mu(s)$, $m(s)$. The equations (\ref{12r}) contain Minkowski force components, one acting along the world line (the tangent, or parallel component) and the other (orthogonal) acting perpendicularly to the world line: 
\begin{eqnarray} 
{\bf U} (dm/ds)={\bf K}_\parallel \ , \   m(d{\bf U}/ds)={\bf K}_\perp, \ & ({\bf K}_\parallel \cdot {\bf K}_\perp) =0 
\label{14r}
\end{eqnarray}
where \  $U^\mu (dm/ds)=K_\parallel^\mu= -U^\mu {\partial W}/{\partial s}$, \ \ $dm/ds=K^s=K_\parallel$ ,  \ \  $m (dU^\mu/ds)= K_\perp^\mu =-W (dU^\mu/ds)$.
So the equations (\ref{12r}) can be expressed in the convenient 4-component form
\begin{eqnarray} 
d(m {\bf U})ds=d{\bf P}/ds={\bf K}_\parallel +{\bf K}_\perp ={\bf K}
\label{15r}
\end{eqnarray}
or
\begin{eqnarray} 
d(mU^\mu)ds=dP^\mu/ds=K^\mu
\label{16r}
\end{eqnarray}
Next stage would be a change of variables and rewriting the equations in terms of $X^i,\ t$ and ``ordinary'' forces $F^i$.
Under conditions of proper mass constancy, the orthogonal component of the Minkowski force vanishes, and (\ref{14r}) reduces to equations of conventional Relativistic Dynamics. The difference due to $dm/ds$ term in (\ref{16r}) may be appreciable in a strong field domain, which is hard to investigate experimentally. 

In our view, the mass-time complementarity concept provides new opportunities of field quantization in the whole energy range. Starting with particle-particle interaction in terms of potential fields, a problem formulations with a field dependent proper mass in terms of wave equations (for example, Klein-Gordon equation) would be an alternative approach to quantum field theories. At present, they encounter different problems, first of all, high-energy divergence and non-renormalizibility (gravitational problem). By exploring the role of proper mass variablity, one may gain a new insight into those problems.  

 %Those who are interested in more details, may see Authors works \cite{Van1},\cite{Van2} with references.  

\section{Conclusion}

To the conclusion, we would like again to emphasize the aspect of SRT physical foundations which was dominant in the paper: unity of mass and time. A relativistic approach to Mechanics requires to work with structureless, that is point particles, nevertheless, having an inner property, a mass.
The origin and relativistic nature of ``mass'' is a matter of future theories but its fundamental significance is already seen in SRT quantum connections. We call the time interval and the mass (two scalars, which are temporary components of 4-position and 4-momentum vectors) complementary because the time rate is directly related to atomic clock proper mass. Thus, particle-wave duality is present in SRT in the form of the de Broglie wave concept based on three universal physical constants $c_0$, $m_0$, $h$. We explored different consequences of complementarity of Minkowski space using a methodology of imaginary observations of moving particles exchanging standard photons. In particular, a scrupulous analysis of the twin paradox was conducted with the conclusion that twins would agree on an equality of travel time record (a number of atomic clock ``ticks'') while the so-called ``elapsed time'' could be different for each twin (depending on comparison conditions). The conclusion is consistent with Lorentz invariance of the proper time interval.  

It is clear that SRT is a model of non-interacting particles that is, forces are turned off. Introduction of field of forces in Minkowski space is subject to SRT Dynamics development as a step to a field theory. A point particle model leads to classical self-energy divergence and related to it field singularity problems. These issues are definitely out of the scope of the present work. However, it was important, in our view, to show possible consequences of the mass-time complementarity concept in Relativistic Dynamics in general form. We know from observations that an external field may influence an atomic clock rate, and the question arises if the effect can be explained by action of Minkowski force on the proper mass. The conclusion was made that such mechanism could exist and should be investigated.

The main objective of the paper is to show the unity of mass and time and its consequences in Relativistic Theory. Yet, we wanted to explain to general physicists, philosophers and just interested, familiar with Physics, persons a physical meaning of time in rigor, Minkowski terms. 
The paper could also be helpful in maintaining reservations for those who would attempt to refute, reinterpret or ``improve'' SRT Kinematics after Einstein.
%, or alter it on a fuzzy road to the realm of field theories. 

%\end{bibliography}


\begin{thebibliography}{99}


%\bibitem {Landau} L. D. Landau and E M. Lifshitz.
%``The Classical Theory of Fields'', Pergamon Press (1975).
%\bibitem{Test1} G. Saathoff, S. Karpuk, U. Eisenbarth, {\it et al}. Phys. Rev. Lett. {\bf 91}, 190403-1 (2003).

%\bibitem {1} For example: Bryan Walles. ``The Farce of Physics''. Internet-posted book (1994). 2d Conference on Problems of Space and Time in Natural Science. St. Petersburg, Russia (1991).

\bibitem {1} Avi Nelson. ``Reinterpretation of the famous train/embankment experiment of relativity''. Eur. J. Phys. {\bf 24}, p. 379 (2003).  
%\bibitem {Moller} C. Moller. ``The Theory of Relativity''. International series of monographs on physics. Delhi, Oxford University Press (1972).
\bibitem {Misner} C. W. Misner, K. S. Thorne, J. A. Wheeler. ``Gravitation''. W. H. Freedman and Company, San Francisco (1973).
\bibitem {Moller} C. Moller. ``The Theory of Relativity''. International series of monographs on physics. Delhi, Oxford University Press (1972).
\bibitem{Test1} G. Saathoff, S. Karpuk, U. Eisenbarth, {\it et al}. Phys. Rev. Lett. {\bf 91}, 190403-1 (2003).
%\bibitem {Misner} C. W. Misner, K. S. Thorne, J. A. Wheeler. ``Gravitation''. W. H. Freedman and Company, San Francisco (1973). 
\bibitem {Pauli} W. Pauli. ``Theory of Relativity'' (translated from German by G. Field). Dover Publication, Inc. New York (1981). Originally in German: ``Relativitatstheorie'', Encyclopedie der matematischen Wissenschafte, V19, B. G. Teubner, Leipzig (1921). 
\bibitem {Penrose} R. Penrose. ``The Road to Reality (A complete guide to the laws of the universe)''. Alfred A. Knopf, New York (2005). 
\bibitem {Tonnelat} Marie-Antoinnette Tonnelat. ``Les Principes de la Theorie et de la Relativite''. Masson Editeurs (1950). 
\bibitem {Sartori} L. Sartori. Understanding Relativity. University of California Press, Berkley, Los Angeles, London (1984).
\bibitem{Synge} J.L.Synge. ``Relativity: The Special Theory''.
North Holland Publishing Company, Amsterdam (1965).
\bibitem {Landau} L. D. Landau and E M. Lifshitz.
``The Classical Theory of Fields'', Pergamon Press (1975).
%\bibitem{Synge} J.L.Synge. ``Relativity: The Special Theory''.
%North Holland Publishing Company, Amsterdam (1965).
\bibitem{Van1} A. Vankov. ``On de Broglie wave nature''. Annales de la foundations Louis de Broglie, Volume 30, 1, p. 15 (2005). 
``Testing relativistic mass-energy concept in Physics of gravity and electricity''. AFLB (Special issue Yang-Mills paper commemoration). Volume 29, 2, p. 1035 (2004). 
\bibitem{Van2} A. Vankov. ``On problem of mass origin and self-energy divergence in Relativistic Mechanics'', arXiv: gr-qc/0311063 (2003). ``Elimination of high-energy divergence in relativistic Lagrangean formulation of particle dynamics''. arXiv: physics/0402117 (2004).



%\bibitem {Goldstein} H. Goldstein. ``Classical Mechanics''. Addison-Wesley Publishing Company; series in Advanced Physics (1950).
%\bibitem {Jackson} J. D.Jackson. ``Classical Electrodynamics'', 3d Edition. John Wiley and Sons, New York (1998).  
%\bibitem {Carlip} S. Carlip. ``Quantum Gravity: a Progress Report.'' arXiv: gr-qc 0108040 (2001). 

%\bibitem{Einstein} A. Einstein. ``Cosmological Considerations on the General Theory of Relativity''. Sitzungsberichte der Preussischen Akad. d. Wiss., (1917), p. 142-152. (1917 translated). 
%\bibitem {Misner} C. W. Misner, K. S. Thorne, J. A. Wheeler. ``Gravitation''. W. H. Freedman and Company, San Francisco (1973). 
%\bibitem {2} C. J. Isham. ``Structural problems facing quantum gravity theory'', {\it Proceedings of the 14th International Conference on General Relativity and Gravitation.} World Scientific, Singapore (1997), 167-209. 
%\bibitem {3} C. Rovelli and L. Smolin. ``The physical Hamiltonian in nonperturbative quantum gravity''. Phys. Rev. Lett., {\bf 72}, (4), (1994), 446-449. 
%\bibitem{4} Robertson H. P., Noonan T. W. ``Relativity and Cosmology''. W. B. Saunders Company (1968).
%\bibitem {Synge} J. L. Synge. ``Relativity: The Special Theory''.
% North Holland Publishing Company, Amsterdam (1965). ``Relativity: The General Theory''.
 %North Holland Publishing Company, Amsterdam (1964). 
%\bibitem {Taylor} E. F. Taylor, J. A. Wheeler. ``Exploring Black Holes''.
%(Introduction to General Relativity Theory). Addison Wesley Longman (2000).
%\bibitem {6} H. Goldstein. ``Classical Mechanics''. Addison-Wesley Publishing Company; series in Advanced Physics (1950). 
%\bibitem{Weinberg} S. Weinberg. Gravitation and Cosmology:
%Principles and Applications of the General Relativity Theory.
%New York, Willey and Sons, Inc. (1972).
%\bibitem{Friedmann} A. Friedmann. ``On the Curvature of Space''. Zeitschrift f\"ur Physik, {\bf 10}, 377-386 (1922).  A. Friedmann. ``On the Possibility of the World with Constant Negative Curvature''. Zeitschrift f\"ur Physik, {\bf 21}, 326-332 (1924) (translated).
%\bibitem{Sitter} W. de Sitter. ``On the Einstein's Theory of Gravitation and its Astronomical Consequences''. Monthly Notices Royal Astronomical Society, {\bf 78}, 3-28 (1917).
%\bibitem {Lemaitre} G. Lamaitre. ``A Homogeneous Universe of Constant Mass and Increasing Radius''. Monthly Notices Royal Royal Astronomical Society. {\bf 91}, 483-490 (1931).
%\bibitem{WalkerRobertson}  H. P. Robertson. ``On the Foundations of Relativistic Cosmology''. Proceedings of the National Academy of Science, {\bf 15}, 822-829 (1929). 
%\bibitem {Peebles} P. J. Peebles. ``Principles of Physical Cosmology''. Princeton University Press (1993).
%\bibitem {Vankov} A. Vankov. ``On the de Broglie Wave Nature''. Annales de la Fondation Louis de Broglie. (To be published).  A. Vankov. Annales de la Fondation Louis de Broglie. ``Testing Relativistic Mass-Energy Concept in Physics of Gravity and Electricity''. Annales de la Fondation Louis de Broglie (the Special Issue dedicated to the 50th anniversary of the Yang-Mills paper, 1954). Volume {\it 29}, 2 (2004).
%\bibitem {Isham} C. J. Isham. ``Structural problems facing quantum gravity theory'', {\it Proceedings of the 14th International Conference on General Relativity and Gravitation.} World Scientific, Singapore (1997), 167-209. 
%\bibitem {Rovelli} C. Rovelli and L. Smolin. ``The physical Hamiltonian in nonperturbative quantum gravity''. Phys. Rev. Lett., {\bf 72}, (4), (1994), 446-449. 
%\bibitem {Landau} L.D.Landau and E.M.Lifshitz.
%``The Classical Theory of Fields'', Pergamon Press (1975).
%\bibitem {Okun} L. B. Okun, K. G. Selivanov, and V. L. Telegdi.  ``On the interpretation of the redshift in a static gravitational field''. Am. J. Phys. {\bf 68}, (2), 115-119. February (2000). S. I. Blinnikov, L. B. Okun. M. I. Vysotsky. ``Critical velocities in general theory of relativity''. arXiv:gr-qc/0310020 (2003).
%\bibitem {Robertson} Robertson H. P., T. W. Noonan ``Relativity and Cosmology''. W. B. Saunders Company (1968).
%\bibitem {Clifford} C. M. Clifford. ``Theory and Experiment in Gravitational Physics''. Cambridge (1993).
%\bibitem {Ashley} N. Ashley. Relativity and the Global Positioning System. Physics Today, May (2002).
%\bibitem{Fishman} G. J. Fishman, P. N. Blatt, R. Malozzi, et. al. Discovery of intense gamma-ray  flashes of atmospheric origin. Science, {\bf 264}, 1313 (1994).



%\bibitem {Vankov2} A. Vankov. ``On an Alternative Cosmology''. arXiv:astro-ph/9811165 (1998);  A. Vankov. ``On matter-antimatter separation in open relativistic material system''. arXiv:astro-ph/9906279 (1999); A. Vankov.  ``On the cosmological aspects of observed high energy cosmic phenomena''.  arxiv:astro-ph/9904045 (1999, submitted to the ANS Centennial Meeting, March 20 - 26, 1999, Atlanta, Georgia); A. Vankov. ``On neutrino family problem''. arXiv:hep-ph/0012036 (2000).
%\bibitem {Danby} G. Danby, J-L. Gaillard, K. Gouliakos, et al. ``Observations of high-energy neutrino reactions and the existence of two kinds of neutrinos''. Phys. Rev. Lett. {\bf 9}, 36-44 (1962).   



%%%%%%%%%%%%%%%%%%%%%%%%%%%%%%%%%%%%%%%
%\bibitem{1} E.F.Taylor, J.A.Wheeler. ``Exploring Black Holes''
% (Introduction to General Relativity Theory). Addison Wesley Longman (2000)

%\bibitem{2} C.W.Misner, K.S.Thorn, J.A.Wheeler. ``Gravitation''.
% W.H.Freeman and Company, San Francisco (1973).

%\bibitem{2a} M. H. Goroff and A. Sagati. Nucl. Phys., B 266, 709 (1986).

%\bibitem{3} J.L.Synge. ``Relativity: The Special Theory''.
% North Holland Publishing Company, Amsterdam (1965).

%\bibitem{4} C.Moller. ``The Theory of Relativity''.
% Oxford University Press (1982).

%\bibitem{5} A.Pais. ``The Science and the Life of Albert Einstein''.
% Clarendon Press, Oxford University Press (1982).

%\bibitem{6} C.M.Clifford. ``Theory and Experiment in Gravitational Physics''.
% Cambridge (1993).

%\bibitem{7} L.B.Okun, K.G.Selivanov, and V.L.Telegdi
% ``On the interpretation of the redshift in a static gravitational field''.
% Am.J.Phys. 68 (2), 115-119. February 2000.

%\bibitem{8} N.Ashley.Relativity and the Global Positioning System.
% Physics Today, May 2002.

%\bibitem{9} G.J.Fishman, P.N.Blatt, R.Malozzi, et. al.
% Discovery of intense gamma-ray  flashes of atmospheric origin.
% Science, 264, 1313 (1994).

%\bibitem{10} L.D.Landau and E.M.Lifshitz.
% ``The Classical Theory of Fields'', Pergamon Press (1975).

%\bibitem{11} S.Weinberg. Gravitation and Cosmology:
% Principles and Applications of the General Relativity Theory.
% John Willey and Sons, Inc. (1972).

%\bibitem{12} J.L.Synge. ``Relativity: the General Theory''.
% North Holland Publishing Company, Amsterdam (1964).

%\bibitem{12a} S.N.Gupta. ``Theory of longitudinal photons in quantum
% electrodynamics''. Proc.Phys.Soc. (London), A 64, 681 (1950).

%\bibitem{12b} K.Bleuler. ``Eine neue methode zur behandlung der longetudinalen
% und scalaren photonen''. Helv.Hys.Acta, 23, 567 (1950).


%\bibitem{13} Louis de Broglie. ``Non Linear Wave Mechanics (A Casual
% Interpretation). Translated by A.Knodel and J.Miller. Elsevier Publishing
% Company (1960).

%\bibitem{14} A.Einstein, B.Podolski, and N.Rosen.
% Phys.Rev., 41, p.777 (1935).

%\bibitem{15} D.Bohm. Quantum Theory. Dover, New York (1957).

%\bibitem{16} J.Bell. Rev.Modern Phys., 38, p.447 (1966).

%\bibitem{weak} S.L.Glashow. Nucl. Phys. 22, 597 (1961); S.Weinberg, Phys. Rev. Lett. 19, 1264 (1967); A.Salam. Proc. of the 8th Nobel Symposium on {\it Elementary particle theory, relativity groups and analiticity}, edited by Svartholm (1969).

%\bibitem{18} W.Heisenberg. ``Introduction to the unified field theory of
% elementary particles''. Max-Planck Institute f\"ur Physik und Astrophysik.
% Interscience Publisher, London, New York, Sydney (1966).

%\bibitem{19} W.Heisenberg. Rev.Mod.Phys., 29, 269 (1957).

%\bibitem{20} W.Heisenberg. Zs.F\"ur Phys., 144, 1, (1956).


\end{thebibliography}
\end{document}